\documentclass[twocolumn]{aastex631}

\usepackage{graphicx}	
\usepackage{amsmath}	
\usepackage{amssymb}	
\usepackage{threeparttable}	
\usepackage{mathtools}

\newcommand{\ssst}{\scriptscriptstyle}
\newcommand{\E}[1]{\times 10^{#1}}

\newcommand{\s}{\,{\rm s}}      
\newcommand{\ps}{\,{\rm s}^{-1}}

\newcommand{\cm}{\,{\rm cm}}    
\newcommand{\km}{\,{\rm km}}
\newcommand{\kms}{$\km\ps$}

\newcommand{\pc}{\,{\rm pc}}
\newcommand{\erg}{\,{\rm erg}}  
\newcommand{\K}{\,{\rm K}}

\newcommand{\um}{\,\mu\rm m}

\newcommand{\Tmb}{T_{\rm mb}}

\newcommand{\Tex}{T_{\rm ex}}

\newcommand{\nH}{n_{\ssst\rm H}}

\newcommand{\VLSR}{V_{\ssst\rm LSR}}

\newcommand{\Spitzer}{{\sl Spitzer}}

\newcommand{\du}{d_{8}}

\newcommand{\snr}{W49B}
\newcommand{\twCO}{$^{12}$CO}   
\newcommand{\thCO}{$^{13}$CO}
\newcommand{\HCOp}{HCO$^+$}   
\newcommand{\NHCOp}{N({\rm HCO^+})}

\newcommand{\Jotz}{$J$=1--0}    
\newcommand{\Jtto}{$J$=2--1}

\graphicspath{{./}{figures/}}

\begin{document}

\title{Unusually high \HCOp/CO ratios in and outside supernova remnant W49B}

\author{Ping Zhou}
\affil{School of Astronomy \& Space Science, Nanjing University, 163 Xianlin Avenue, Nanjing~210023, China}
\affil{Key Laboratory of Modern Astronomy and Astrophysics, Nanjing University, Ministry of Education, Nanjing~210023, China}
\email{pingzhou@nju.edu.cn}

\author{Gao-Yuan Zhang}
\affil{Departamento de Astronomía, Universidad de Concepción,
Concepción, Chile}
\email{zgy0106@gmail.com}

\author{Xin Zhou}
\affil{Purple Mountain Observatory and Key Laboratory of Radio Astronomy, Chinese Academy of Sciences, 10 Yuanhua Road, Nanjing 210023, China}

\author{Maria Arias}
\affil{Leiden Observatory, Leiden University, PO Box 9513, 2300 RA, Leiden, The Netherlands}

\author{Bon-Chul Koo}
\affil{Department of Physics and Astronomy, Seoul National University, Seoul 08826, Republic of Korea}

\author{Jacco Vink}
\affil{Anton Pannekoek Institute for Astronomy, University of Amsterdam, Science Park 904, 1098 XH Amsterdam, The Netherlands}
\affil{GRAPPA, University of Amsterdam, Science Park 904, 1098 XH Amsterdam, The Netherlands}
\affil{SRON, Netherlands Institute for Space Research, Sorbonnelaan 2, 3584 CA Utrecht, The Netherlands}

\author{Zhi-Yu Zhang}
\affil{School of Astronomy \& Space Science, Nanjing University, 163 Xianlin Avenue, Nanjing~210023, China}
\affil{Key Laboratory of Modern Astronomy and Astrophysics, Nanjing University, Ministry of Education, Nanjing~210023, China}

\author{Lei Sun}
\affil{Anton Pannekoek Institute for Astronomy, University of Amsterdam, Science Park 904, 1098 XH Amsterdam, The Netherlands}
\affil{School of Astronomy \& Space Science, Nanjing University, 163 Xianlin Avenue, Nanjing~210023, China}

\author{Fu-Jun Du}
\affil{Purple Mountain Observatory and Key Laboratory of Radio Astronomy, Chinese Academy of Sciences, 10 Yuanhua Road, Nanjing 210023, China}

\author{Hui Zhu}
\affil{Key Laboratory of Optical Astronomy, National Astronomical Observatories, Chinese Academy of Sciences, Beijing 100012, China}

\author{Yang Chen}
\affil{School of Astronomy \& Space Science, Nanjing University, 163 Xianlin Avenue, Nanjing~210023, China}
\affil{Key Laboratory of Modern Astronomy and Astrophysics, Nanjing University, Ministry of Education, Nanjing~210023, China}

\author{Stefano Bovino}
\affil{Departamento de Astronomía, Universidad de Concepción,
Concepción, Chile}

\author{Yong-Hyun Lee}
\affil{Samsung SDS, Olympic-ro 35-gil 125, Seoul, Republic of Korea}

\begin{abstract}

Galactic supernova remnants (SNRs) and their environments provide the nearest laboratories to study SN feedback. 
We performed molecular observations toward SNR \snr, the most luminous Galactic SNR in the X-ray band, aiming to explore signs of multiple feedback channels of SNRs on nearby molecular clouds (MCs).
We found very broad \HCOp\ lines with widths of $dv\sim 48$--75~\kms\ in the SNR southwest, providing strong evidence that \snr\ is perturbing MCs at a systemic velocity of $\VLSR=61$--$65~\km\ps$, and placing \snr\ at a distance of $7.9\pm 0.6$~kpc.
We observed unusually high-intensity ratios of \HCOp~\Jotz/CO~\Jotz\ not only at shocked regions ($1.1\pm 0.4$ and $0.70\pm 0.16$), 
but also in quiescent clouds over 1~pc away from the SNR's eastern boundary ($\ge 0.2$).
By comparing with the magnetohydrodynamics (MHD) shock models, we interpret that
the high ratio in the broad-line regions can result from a cosmic-ray (CR) induced chemistry in shocked MCs, where the CR ionization rate is enhanced to around 10--$10^2$ times of the Galactic level.
The high \HCOp/CO ratio outside the SNR is probably caused by the radiation precursor, while the luminous X-ray emission of \snr\ can explain a few properties in this region.
The above results provide observational evidence that SNRs can strongly influence the molecular chemistry in and outside the shock boundary via their shocks, CRs, and radiation.
We propose that the \HCOp/CO ratio is a potentially useful tool to probe an SNR's multichannel influence on MCs.

\end{abstract}


\section{Introduction} \label{sec:intro}

Supernovae (SNe) are among the most energetic explosions in galaxies. SNe quickly fade away but leave their energy and metals in their supernova remnants (SNRs).
While expanding in the interstellar medium (ISM),
SNRs return a huge amount of energy (typically around the order of $10^{51}~\erg$) to the galaxies via shocks, radiation, and cosmic rays  \citep[CRs;][]{vink20}.

Theoretical studies show that SN feedback strongly regulates star formation and galaxy evolution \citep[e.g.,][]{scannapieco08,ostriker11}. 
Nevertheless, there are large uncertainties on the level of SN feedback, because it relies on the understanding of SN population and energy \citep{keller22}, and a clear knowledge of how multiple feedback channels (shocks, radiation, CRs) of SNR influence the ISM.
Observing SNRs is a direct way to quantify the SN feedback and test those SN parameters applied in numerical simulations. For example, \cite{koo20} used HI and molecular observations of seven SNRs and found that their momentum and kinetic energy agree well with those used in numerical simulations for typical SNe.
In addition to the mechanical output, SNRs release part of their energy as CRs and radiation, which
can influence the preshock area \citep[see e.g.,][]{vaupre14}.

As a relatively dense phase of the interstellar gas, molecular clouds (MCs) can efficiently slow down the SNR shocks, absorb the radiation, and interact with CRs.
While shocks have been the most widely studied heating channel from SNRs, it is also crucial to learn the influence of SNRs' CRs and radiation.
CRs are the main heating sources of dense molecular gas where UV emission is shielded \citep{goldsmith78}.
X-ray emission has been regarded as important ionization and heating source of neutral gas in active galactic nuclei (AGN) and primitive galaxies \citep{maloney96, lebouteiller17, vallini19}.
The term X-ray dissociation regions or X-ray dominated regions (XDRs) has been used to describe the regions where the heating and chemistry of the gas are controlled mainly by the intense X-ray fluxes \citep{maloney96}.
Due to the small cross section (decreasing with X-ray energy),
the X-ray photons can traverse deep into
the MCs and irradiate the gas opaque to UV photons.
So far, more than 70 SNRs have been found to interact (or likely interact) with MCs \citep[see][and references therein]{frail96,reach06,chen14,lee19}. Many of them are bright in the X-ray and $\gamma$-ray bands \citep{slane15}.
They also have large infrared-to-X-ray flux ratios suggesting a radiative heating of dust \citep{koo16}. 
These SNRs and their environments provide the nearest laboratories to study the materials exposed to highly enhanced CRs and X-ray radiation.

SNR W49B is the most luminous Galactic SNR in the X-ray band \citep{immler05} and has been subject of 
substantial X-ray studies on its plasma properties and SNR type 
\citep{hwang00,miceli06,keohane07,ozawa09, zhoux11,lopez13a,lopez13b,zhou18a,yamaguchi18,zhang19,sun20,hollandashford20,siegel20}.
\snr\ is detected in H.E.S.S. and Ferm-LAT observations, which suggest its GeV--TeV $\gamma$-ray emission 
is produced through the decay of $\pi^0$ created by the interaction between accelerated CR protons and nearby dense materials
\citep{hess18}. 
Both the X-ray and $\gamma$-ray studies suggest that \snr\ is evolving in a dense medium.
The IR study of \snr\ shows that the dense gas and dust have multiple temperature components and the SNR is evolving in a cloudy environment \citep{keohane07,zhu14}.

Although \snr\ appears to be a good target to study SNR feedback onto dense gas, 
there are disputes on the SNR's distance and associated clouds.
Previous HI emission and absorption studies have not reached a consensus on the SNR distance issue.
While some HI studies claimed that \snr\ lies 3--4~kpc closer than  the star-forming region W49A to the west  at $\sim 11$~kpc \citep{kazes70,radhakrishnan72,zhang13}, others preferred a distance of 11--14~kpc \citep{lockhard78,brogan01}. 
The \twCO\ observation shows a morphological relation between
the molecular gas and the SNR at the local-standard-of-rest velocity ($\VLSR$) at $\sim 40~\km\ps$, corresponding to a distance of 9--10~kpc \citep{chen14,zhu14}.
The first kinetic evidence of \snr--MC association was provided by a recent study by \cite{lee20}, who found
that the near-infrared H$_2$ line centroid is at $\VLSR= 64\pm2~\km\ps$ and proposed an SNR distance of $\sim 7.5$~kpc.
This velocity is
consistent with those suggested by the earlier HI absorption studies \citep[e.g.,][]{kazes70}
and radio recombination lines
from ionized gas toward \snr\ \citep{downes74,liu19}.
The dispute on \snr's distance is not ended. Recently, \cite{sano21} proposed a distance of 11~kpc for \snr\ using \twCO\ observations, which reveal warm MCs at $\VLSR\sim 10~\km\ps$.

Motivated by the above issues,
we performed dedicated molecular observations toward \snr\ at the millimeter wavelength, aiming to 
achieve two goals: (1) to search for MCs perturbed by this SNR, and thus pin down the SNR distance; 
(2) more importantly, to explore signs of multiple feedback channels of an SNR on nearby MCs.
We found unusually high \HCOp/CO ratios in and outside \snr, which are signs
that the CRs, shocks, and radiation of \snr\ strongly influence the chemistry of the environmental gas.
We present our observations and data in Section~\ref{sec:obs}. The data analysis is shown in Section~\ref{sec:results}. 
Section~\ref{sec:discussion} shows our 
interpretation of the unusual chemical properties, with discussions of relevant issues.
The conclusion is summarized in Section~\ref{sec:conclusion}.

\begin{figure*}
	\includegraphics[width=0.32\textwidth]{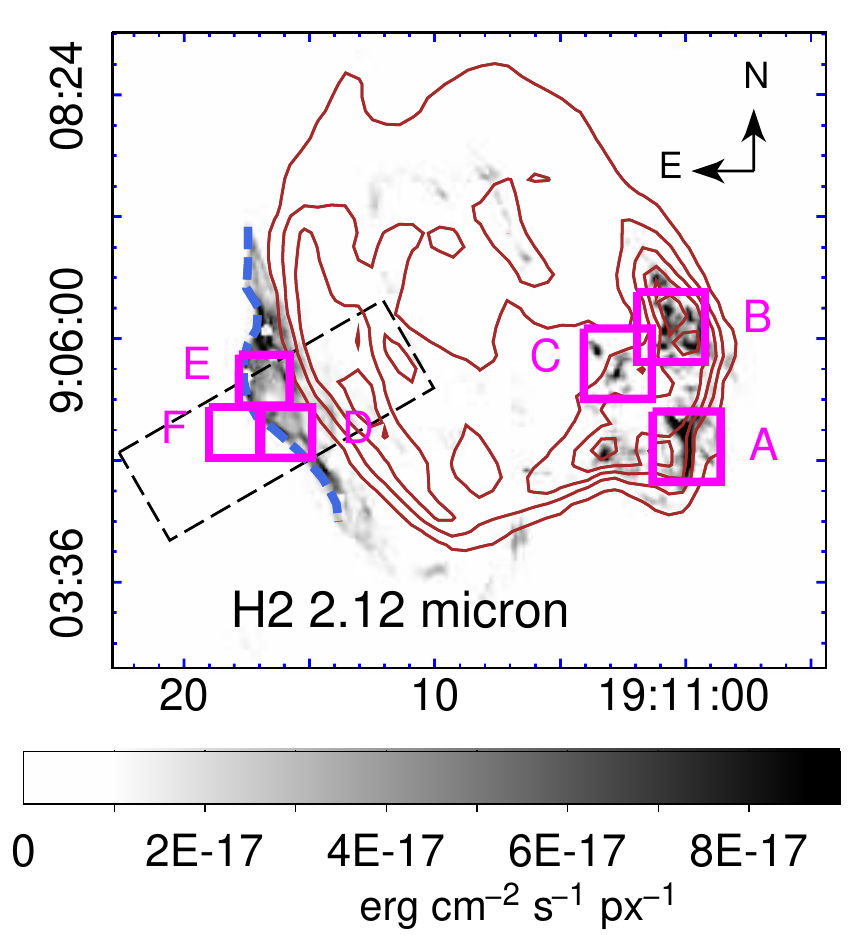}
	\includegraphics[width=0.32\textwidth]{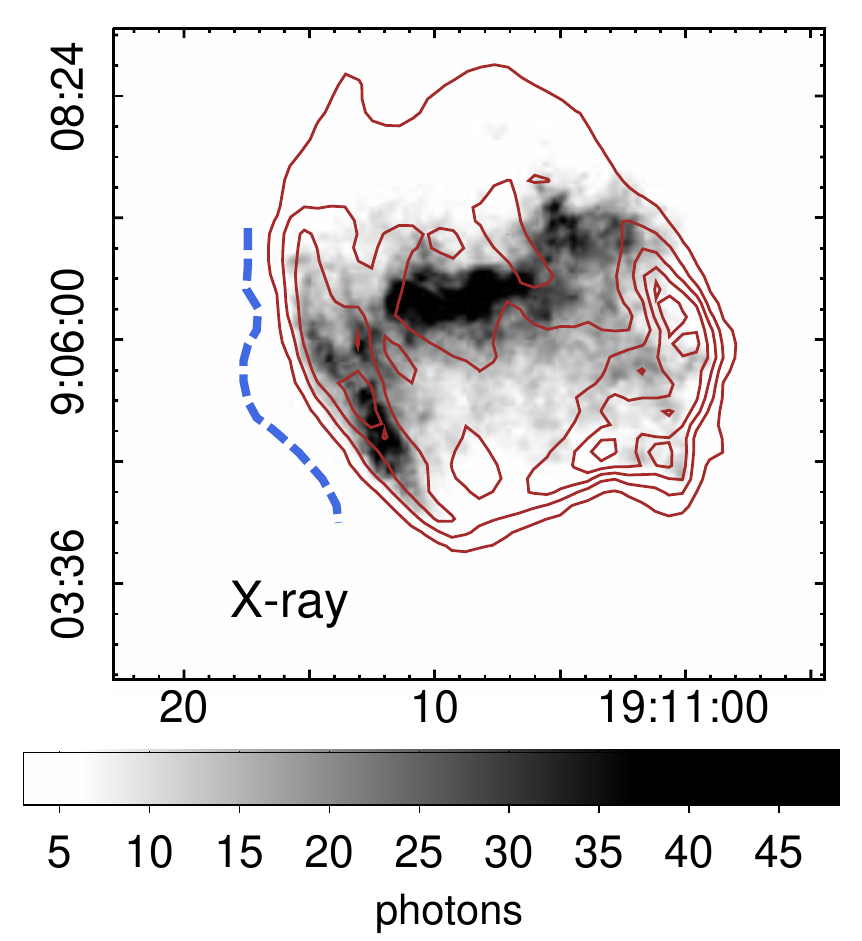}
	\includegraphics[width=0.32\textwidth]{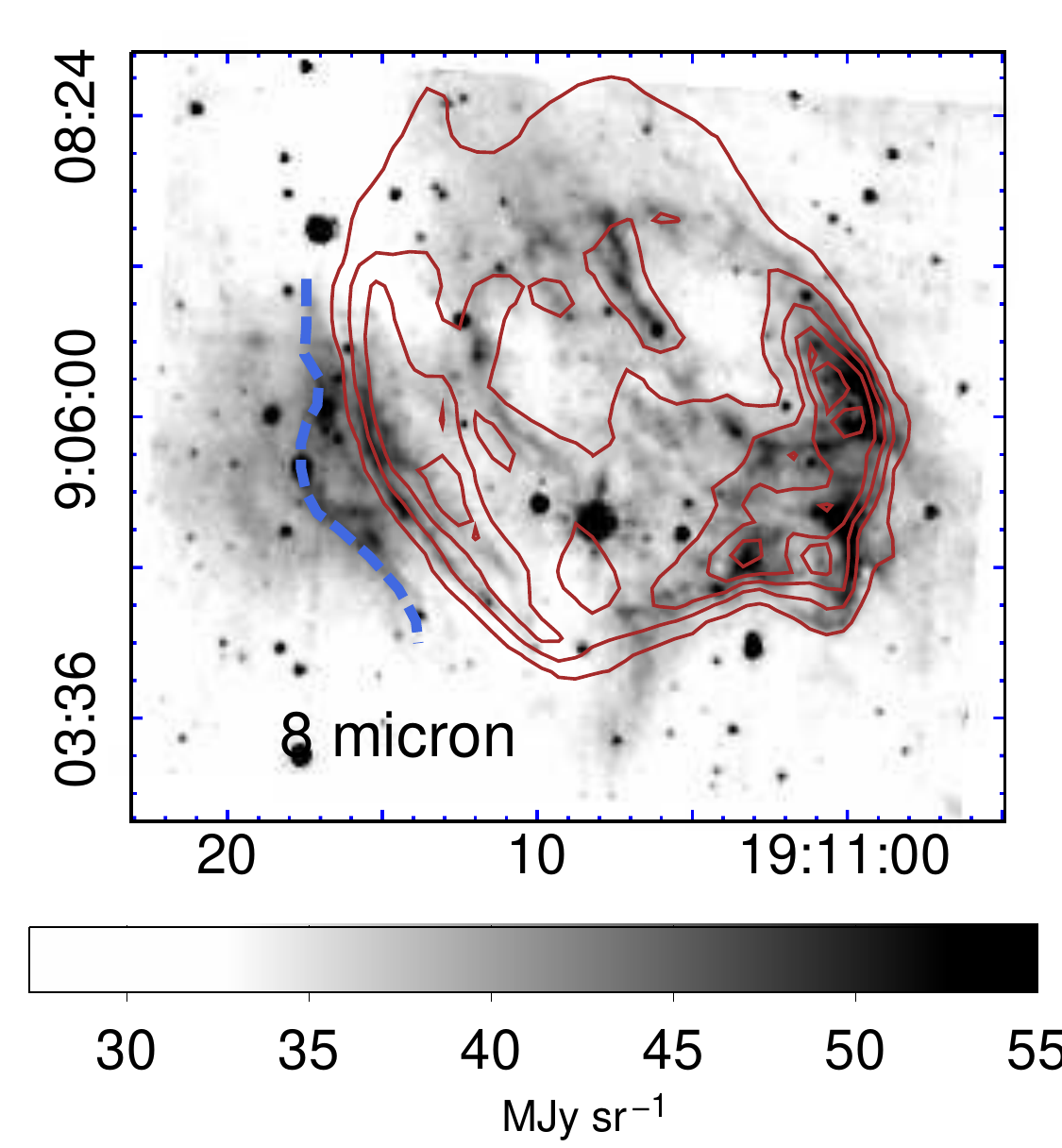}
    \caption{
    2.12~$\um$ H$_2$ \citep[left panel,][]{lee19}, Chandra 0.3--10~keV X-ray photon image \citep[middle panel;][]{zhou18a}, and
    Spitzer 8~$\um$ image (right panel)
    of \snr\ with the VLA 327~MHz radio contours overlaid. 
    The J2000 equatorial coordinates are used.
    The six square boxes in the first panel denote
    the regions for molecular line extraction as shown in Figure~\ref{fig:spec}.
    The black, dashed region in the left panel is selected to compare brightness profiles of different wavelengths (see Figure~\ref{fig:cutline}).
    The blue, dashed curve in the SNR east delineates the eastern boundary of the H$_2$ filament.
	}
    \label{fig:4band}
\end{figure*}

\section{observation and data} \label{sec:obs}

We performed millimeter observations toward \snr\ using the IRAM 30 m telescope during 2019 May 15--16 and 2020 July 24--27 (Project IDs: 167-18 and 024-20; PI: P.\ Zhou).
We used the Eight MIxer Receiver (EMIR) to observe
two frequencies (3~mm and 1~mm or 3~mm and 2~mm) 
simultaneously in the position-switching model.
On-the-fly and on-off observing modes were applied for mapping
and pointing observations, respectively.
The back end of fast Fourier transform spectrometers (FTSs) provided a frequency resolution of 195 kHz,
which corresponds to a velocity resolution of $\sim 0.25$~\kms\ at 235~GHz.
The half-power beamwidth (HPBW) of the telescope 
was $10\farcs{7}$ at 230 GHz, $20\farcs{3}$ at 115 GHz,
and $29''$ at 86~GHz. The main-beam efficiencies at these three frequencies were 59\%, 78\%, and 81\%, respectively \footnote{http://www.iram.es/IRAMES/mainWiki/Iram30mEfficiencies}.

We mapped \snr\ in the frequency ranges 83.7--91.5~GHz, 109.5--117.3~GHz, and 225.1--236.0~GHz. 
The mapped region is similar to the field of view (FOV)
of the multiband images in Figure~\ref{fig:4band}.
The observations are deeper in the east, west, and south regions to probe faint emission near the SNR boundary.

\begin{table}
	\centering
	\footnotesize
	\caption{Information for the IRAM~30m observation}
	\label{tab:obs}
	\begin{threeparttable}
	\begin{tabular}{lcc} 
	\hline
	\hline
	Line & Frequency  (GHz) &  image rms$^a$ (mK) \\
	\hline
	\twCO~\Jtto  & 230.540 & 28--60  \\
	\twCO~\Jotz & 115.271 & 32--64  \\
	\thCO~\Jotz &  110.201 & 17--46 \\
	HNC~\Jotz & 90.664 & 10--34  \\
	\HCOp~\Jotz & 89.189 &  12--39 \\
	HCN~\Jotz & 88.632 & 12--42 \\
	\hline
	\end{tabular}
    \begin{tablenotes}
        \item[a] The rms range of the data cube. The minimum rms is taken from the eastern and western regions of the SNRs.
    \end{tablenotes}
\end{threeparttable}
\end{table}
	
In this paper, we focus on CO and \HCOp\ lines,  but we also detected a few other lines such as HCN and HNC lines.
The line frequency and rms of a few molecular lines are tabulated in Table~\ref{tab:obs}. All the IRAM~30m data were reduced using the GILDAS software (vers.\ 10oct18\footnote{ http://www.iram.fr/IRAMFR/GILDAS/}).

For comparison purposes, the data cubes of the \HCOp,  \twCO, \thCO, and HCN emission were resampled to have the same velocity resolution of $0.7~\km\ps$, while a larger velocity bin is sometimes used for broad lines to increase the signal-to-noise ratios.
The pixel sizes of \twCO~\Jtto\ and \twCO/\thCO~\Jotz\ are $5''$ and $10''$, respectively.
In the ratio maps, the \twCO~\Jtto\ data are convolved to $30''$ and regridded with a pixel size of $10''$.

\begin{figure*}
	\includegraphics[width=0.49\textwidth]{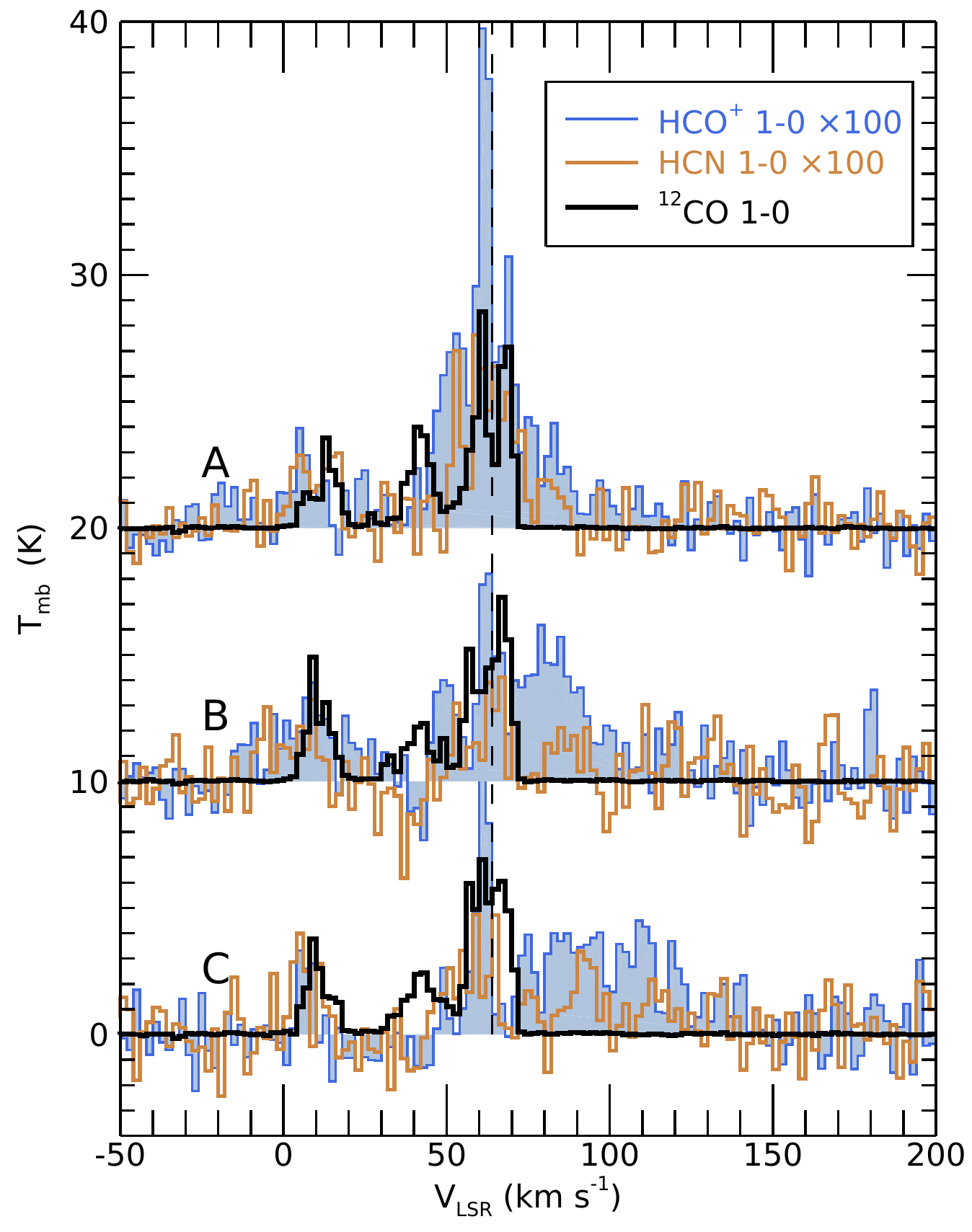}
	\includegraphics[width=0.49\textwidth]{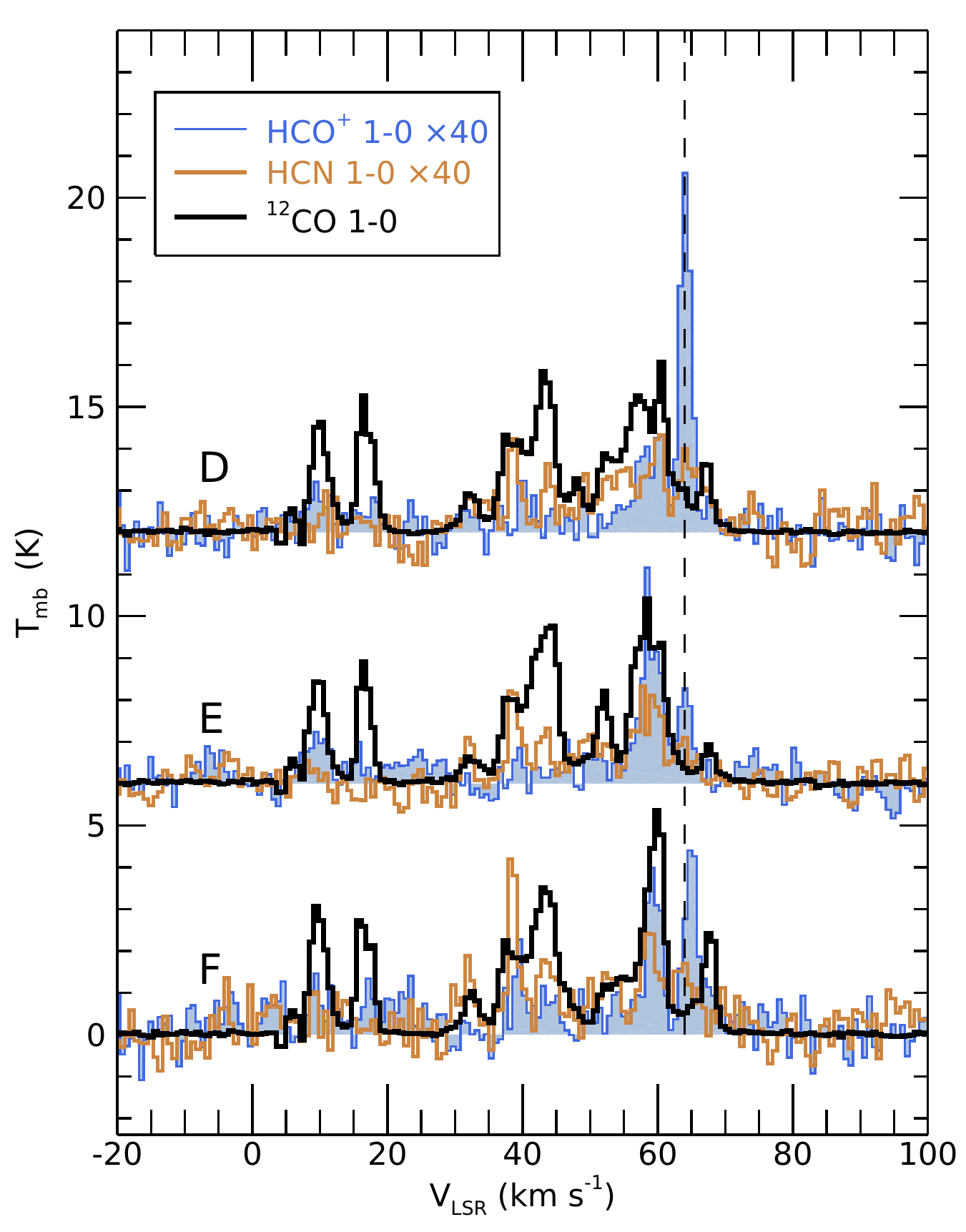}
    \caption{
    Spectra of \twCO\ and scaled \HCOp\ and HCN at six regions labeled in Figure~\ref{fig:4band}. 
    The dashed vertical line indicates the LSR velocity of 64~\kms, where a high \HCOp/CO line ratio is found outside the SNR.
   All data were convolved to reach the same beam size of $30''$.
    \label{fig:spec}}
\end{figure*}

We also used multiwave band maps for comparison
(see Figure~\ref{fig:4band}).
The VLA 327~MHz radio continuum emission data were adopted from \cite{lacey2001}. 
The 2.12$\mu$m H$_2$ image from \cite{lee19}
is reduced 
using the  UKIRT Widefield Infrared Survey for H2 \citep[UWISH2;][]{froebrich11}.
The Chandra X-ray image in 0.3--10~keV was obtained from \cite{zhou18a} using
the observations in 2000 (obs. ID: 117; PI: Stephen Holt) and 2011 (obs. IDs: 13440 and 13441; PI: Laura Lopez). 
We retrieved the 8 $\mu$m IR emission from
the Galactic Legacy Infrared Midplane Survey Extraordinaire \citep[GIMPSE;][]{churchwell09} using the \Spitzer\ telescope.
The angular resolutions of the radio, H$_2$, X-ray, and 
8$\mu$m images are $6\farcs{6}\times 6\farcs{2}$, $0\farcs{7}$ (median seeing), $0\farcs{5}$ (on-axis), and $\sim 2''$, respectively (see aforementioned references for details of each image).

\section{Results} \label{sec:results}

\subsection{Search for shocked molecular gas} \label{sec:broad}

\begin{table*}
	\centering
	\caption{Parameters of the \HCOp\ and HCN emission in six regions}
	\begin{threeparttable}
	\label{tab:lines}
	\begin{tabular}{lcccccc} 
		\hline
		Region$^a$  & \multicolumn{4}{c}{HCO$^+$ (1-0)} & HCN (1-0)  \\ 
		\cline{2-5} 
		& \multicolumn{3}{c}{Gaussian fit} $^b$ \\
		\cline{2-4}
		& $V_{\rm LSR}$ & $\Delta v$  &  $T_{\rm peak}$ & $\int T_{\rm mb} dv$ &
		$\int T_{\rm mb} dv$ \\
		& (km s$^{-1}$)   & (km s$^{-1}$) 
		& (K) & (K~km~s$^{-1}$)  $^c$ &
		(K~km~s$^{-1}$)  $^c$& \\
		\hline
		A (broad) 
		& $61.8\pm 0.8$ & $47.5\pm 2.9$ & $0.064\pm 0.005$
		& $3.07\pm 0.17$  & $1.60\pm 0.15$ 
		\\
		~~~(narrow)
		& $61.9\pm 0.1$ & $5.7\pm 0.4$ & $0.14\pm 0.1$ 
		\\
		~~~(narrow)
		& $69.3\pm 0.2$ & $2.8\pm 0.7$ & $0.06\pm 0.01$
		\\
		B (broad)
		& $77.3\pm 0.9$ & $61.7\pm 3.0$ & $0.044\pm 0. 0.002$
		& $ 2.33\pm 0.13$ & $0.88\pm 0.23$ 
		\\
		~~~(narrow)
		& $62.6\pm 0.2$ & $5.2\pm 0.7$ & $0.061\pm 0.006$
		\\
		C (broad)
		& $97.8\pm 1.9$ & $75.3\pm 6.5$ & $0.035\pm 0.002$
		& $2.45\pm 0.19$ & $1.03\pm 0.19$ 
		\\
		~~~(narrow)
		& $61.4\pm 0.1$ & $5.0\pm 0.4$ & $0.13\pm 0.01$
		\\
		D
		& $63.8\pm 0.3$ & $2.4\pm 0.1$ & $0.192\pm 0.008$
		& $1.00\pm 0.10$ & $1.05\pm 0.10$ 
		\\
		E
		& $63.8\pm 0.2$ & $3.6\pm 0.6$ & $0.054\pm 0.007$
		& $0.82\pm 0.13$ & $0.75\pm 0.09$ 
		\\
		F
		& $64.7\pm 0.1$ & $3.7\pm 0.3$ & $0.11\pm 0.01$
		& $1.31\pm 0.12$ & $1.20 \pm 0.21$ 
		\\
	
		\hline
	\end{tabular}
    \begin{tablenotes}
	    \item[a]{The selected regions are displayed in the upper-left panel of Figure~\ref{fig:4band}, with the \HCOp\ and HCN spectra shown in Figure~\ref{fig:spec}.}  
        \item[b] The \HCOp\ line centroid, FWHM line width and peak brightness fitted
        using one (D--F) or more Gaussian lines (A--C).  
        Only line components with $\VLSR >  60~\km\ps$ are fitted.
        The \HCOp\ spectrum is composed of a ``broad'' and one or two ''narrow'' components at regions A, B, or C, but only one narrow line at regions D, E, or F.
        \item[c] \HCOp\ or HCN intensity integrated for all line components in the range of $\VLSR=40$--150~km s$^{-1}$.
    \end{tablenotes}
\end{threeparttable}
\end{table*}

To search for MCs associated with \snr,  we first 
investigated molecular line profiles across the SNR. 
Finding broadened molecular lines can provide 
strong evidence for a shock perturbation on the MCs \citep[see e.g.,][]{jiang10}, and also helps to determine the SNR distance.

We found very broad \HCOp\ \Jotz\ emission only in the southwestern regions of \snr\ at $\VLSR=40$--150~\kms
(see the spectra in Figure~\ref{fig:spec}.
All the regions correspond to strong H$_2$ emission in the southwest (see the box regions A, B, and C in Figure~\ref{fig:4band}).

By fitting the broad lines with Gaussian profiles, we obtained line widths (FWHM) of $48\pm 3~\km\ps$,  $62\pm 3~\km\ps$, and $75\pm 7~\km\ps$ at regions A, B, and C, respectively (see Table~\ref{tab:lines}).
These lines are too broad to be attributed to quiescent gas but are strong evidence of MCs accelerated by the SNR shock. 
The line widths are also distinctly larger than those predicted in C-type shock, where the H$_2$ molecules survive up to a shock velocity of 45~\kms\ \citep{draine83}.
At region C, only one prominent narrow \HCOp\ line is seen at $\VLSR\sim 61~\km\ps$, while the faint line at $\VLSR\sim 10~\km\ps$ is too far away from the broad component to be related. 
There are also narrow \HCOp\ lines at $\VLSR=61$--63~\kms\ at regions A and B, and at  $\VLSR=64$--65~\kms\ at regions D--F (see further discussions in \S~\ref{sec:spatial} and \S~\ref{sec:discussion}).
We fitted the narrow \HCOp\ lines with $\VLSR > 60~\km\ps$ using Gaussian lines and tabulate the fit results in Table~\ref{tab:lines}.

\begin{figure}
	\includegraphics[width=0.45\textwidth]{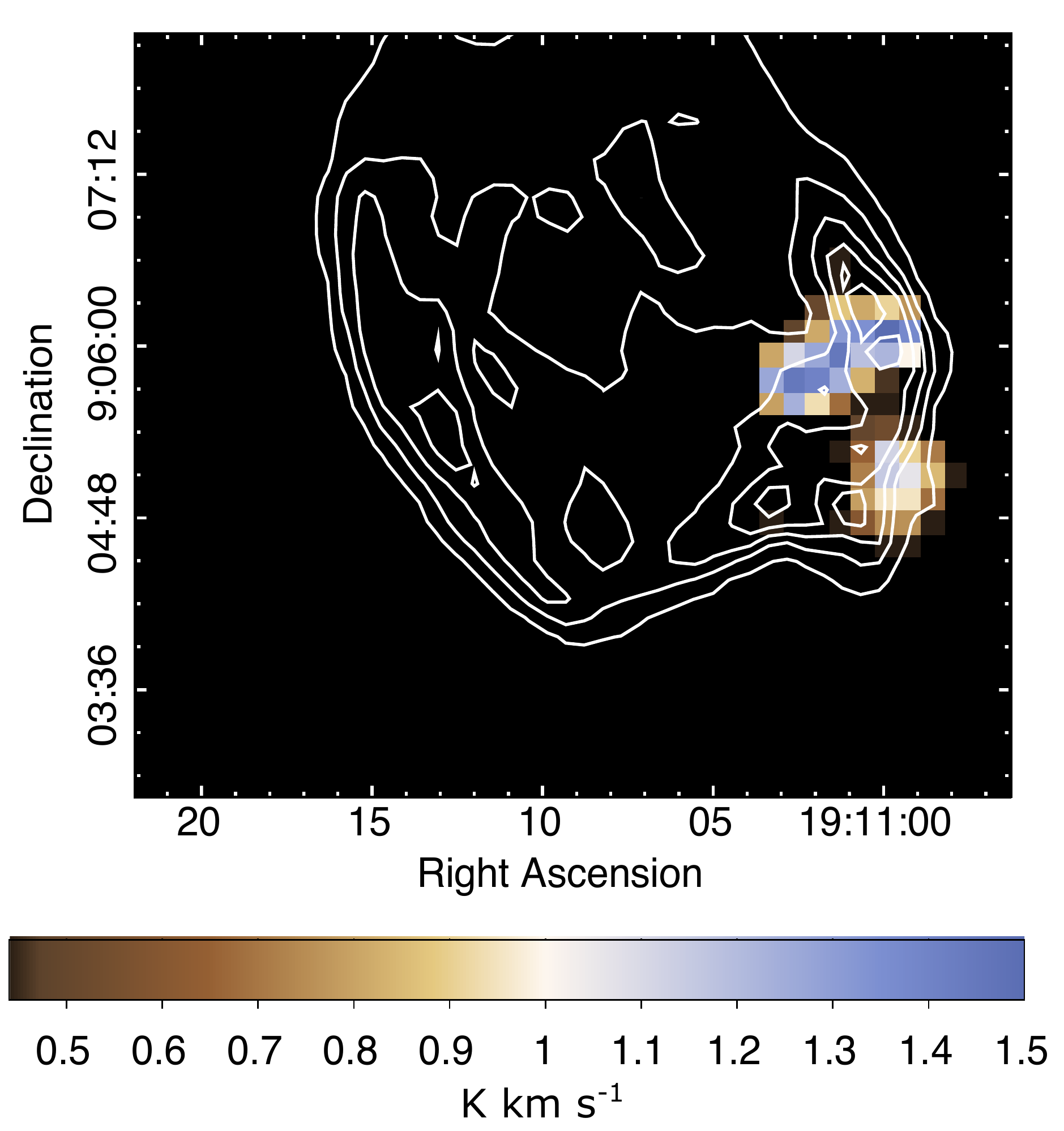}
    \caption{
    Distribution of the shocked \HCOp\ gas in the velocity range $\VLSR=70~\km\ps$ to 120~\kms, overlaid with the 327~MHz radio contours. 
    Values with $\ge 5\sigma$ detection are shown.
    \label{fig:hcop_70_120}}
\end{figure}

Our study supports an SNR systemic velocity of $\VLSR=61$--65~\kms, 
where we found narrow lines (2--$6 \km\ps$ for \HCOp\ emission based on the Gaussian fit; Table~\ref{tab:lines}) nearest to the broad emission and with relatively high \HCOp/CO intensity ratios.
The systemic velocity is consistent with that obtained from the previous H$_2$ spectral study \citep[$64\pm2~\km\ps$,][]{lee20}.
These narrow line components can be explained as
the preshock gas.
This LSR velocity, combined with the measured Galactic rotation curve  \citep{reid14},
places \snr\ at a far distance of $7.9\pm 0.6$~kpc or a near distance of $4.1\pm 0.6$~kpc, where the uncertainty is given  at the 1 $\sigma$ level based on the Monte Carlo method by \cite{wenger18}\footnote{https://www.treywenger.com/kd}.
Due to the HI absorption up to $\VLSR=70~\km\ps$ \citep[the LSR velocity of the tangent point at $\sim 6$~kpc; see e.g.][]{zhu14}, 
\snr\ should be a bright radio source behind the tangent point. Therefore, \snr\ is located at a distance of $7.9\pm 0.6$~kpc.
This means that \snr\ is far away from the star-forming region W49A at a distance of $11.1^{+0.8}_{-0.7}$~kpc \citep[based on parallax measurements using H$_2$O masers;][]{zhang13}.

\begin{figure*}
    \centering
	{\includegraphics[width=0.4\textwidth]{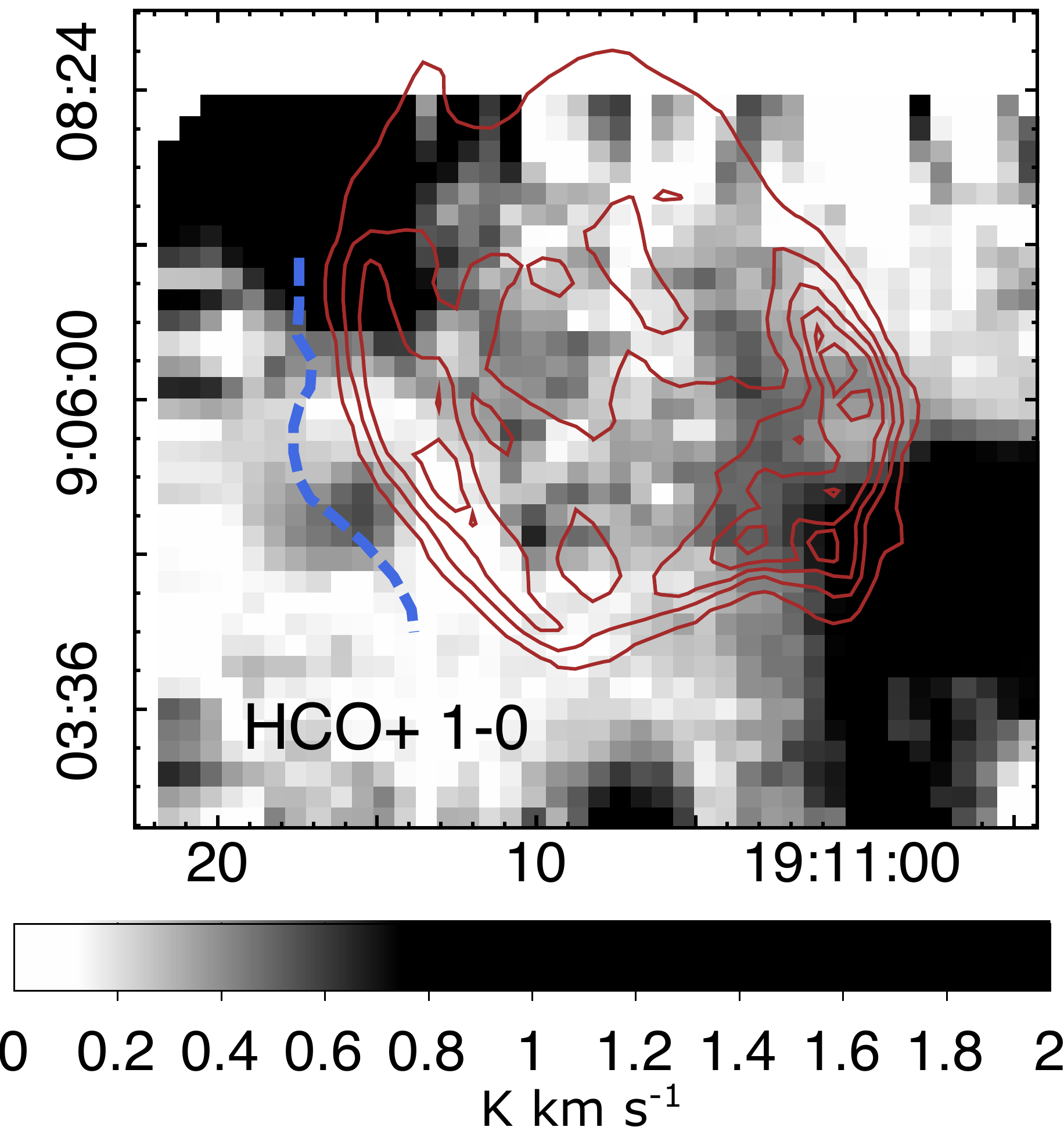}
	\includegraphics[width=0.4\textwidth]{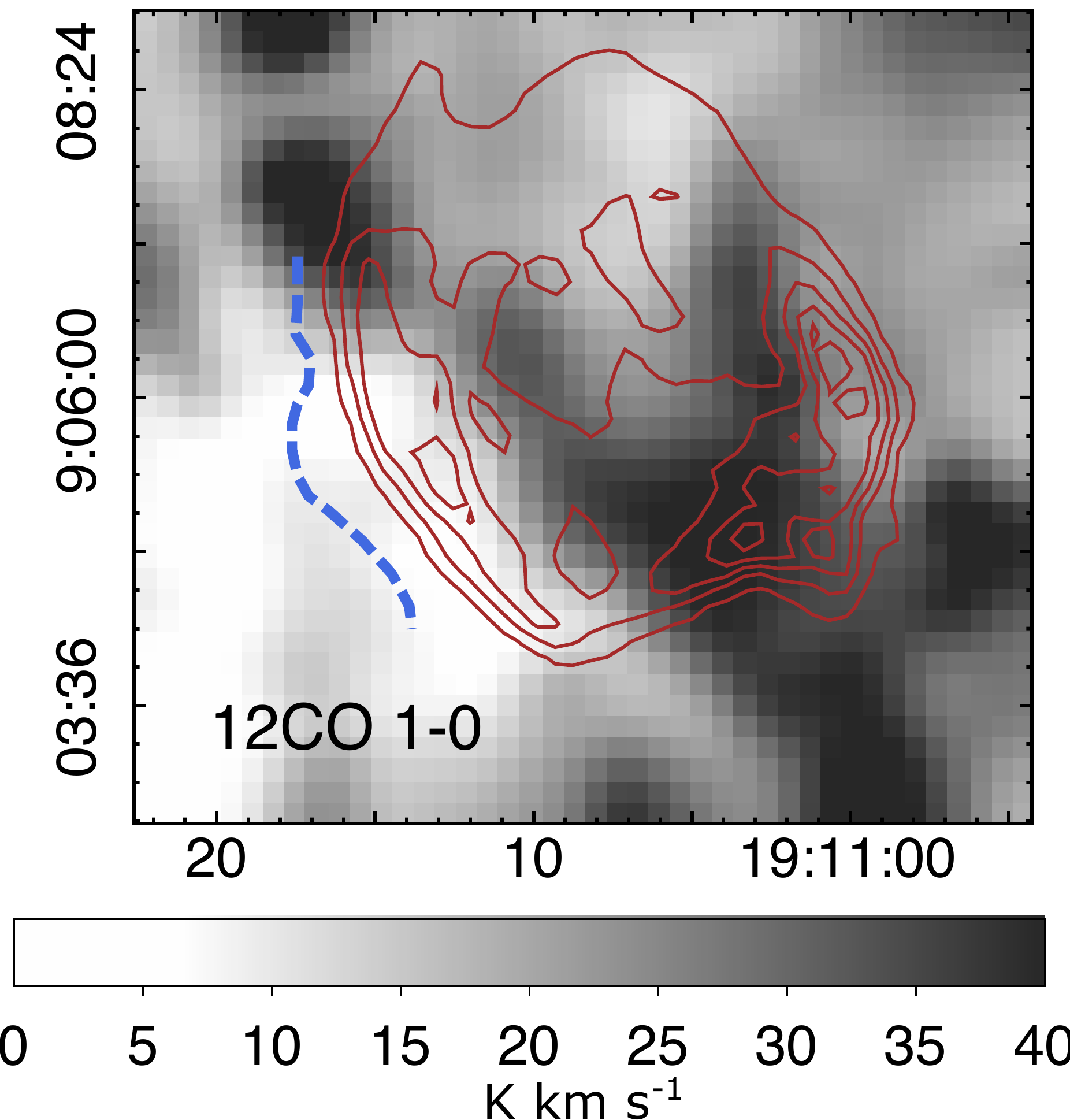}}
	{\includegraphics[width=0.4\textwidth]{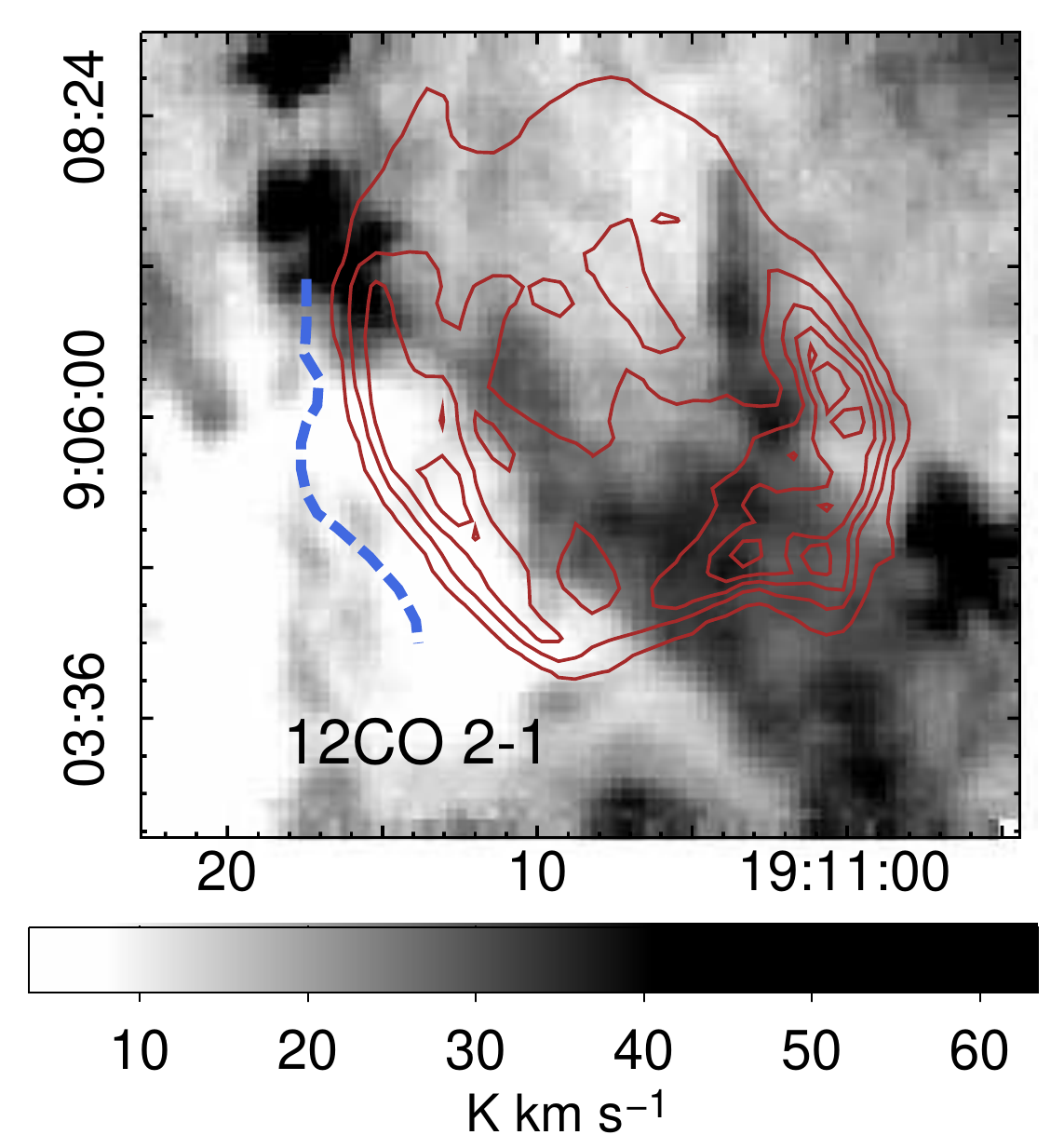}
	\includegraphics[width=0.4\textwidth]{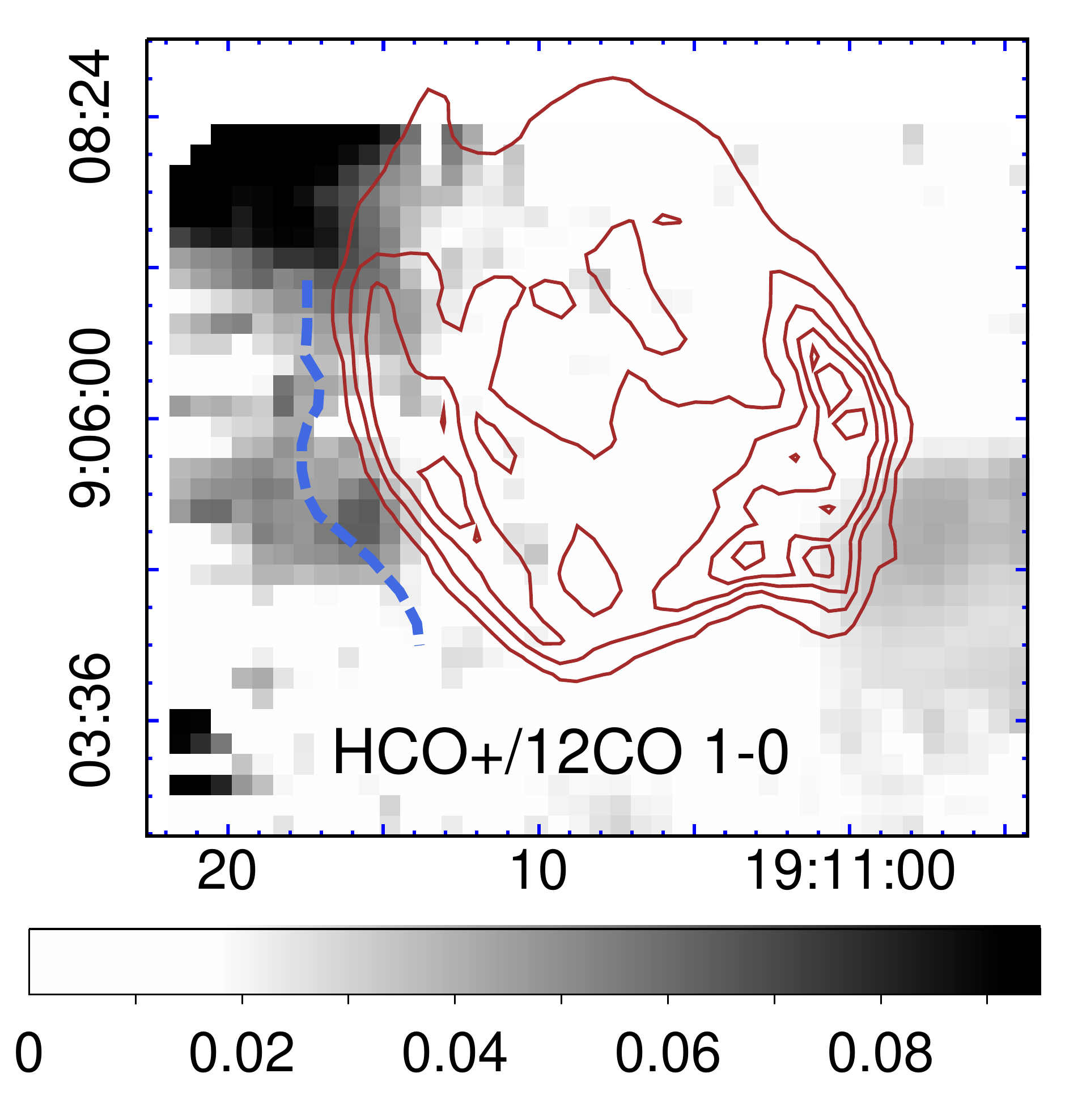}}
    \caption{
    \HCOp\ \Jotz, \twCO\ \Jotz, \twCO~\Jtto, and \HCOp/\twCO\ \Jotz\ images toward \snr\ in $\VLSR=61$--67~\kms\ with the VLA 327~MHz radio contours overlaid (J2000 equatorial coordinates).
    The blue, dashed curve in the SNR east delineates the eastern boundary of the H$_2$ filament.
	}
    \label{fig:hcop_co}
\end{figure*}

A centroid shift of the \HCOp\ broad line is clearly seen from regions A to C, indicating different viewing angles to the shock.
The shift reaches the maximum of $\sim 36~\km\ps$ at region C.
In regions B and C, we only found red-shift \HCOp\ line wings, corresponding to the receding gas that is shocked in the far side of the SNR. 
The MCs at region A are shocked perpendicular to the line of sight, as its \HCOp\ profile shows both red-shift and blue-shift wings,
and  the broad-line centroid is the same as the systemic velocity. 
An interaction with dense MCs explains  the deformed radio morphology in the southwest.
The schematic views of clouds A--F from the observer and the western side are given in Figure~\ref{fig:cartoon} in the appendix.

Figure~\ref{fig:hcop_70_120} shows the spatial distribution of shocked \HCOp\ in the velocity range $\VLSR=70$--120~\kms. 
This velocity range corresponds to the red wing of broadened lines.
We have not found any broad \HCOp\ or \twCO\ lines in other regions of the SNR, except for those visible in Figure~\ref{fig:hcop_70_120}. 
This means that the kinetic evidence of SNR--MC interaction has only been established in the southwestern regions of \snr.
The distribution of \twCO~\Jtto\ and \HCOp~\Jotz\ in the velocity range $\VLSR=0$--$80~\km\ps$ with a step of $5~\km\ps$ are shown in Figures~\ref{fig:12co21_grid} and \ref{fig:hcop_grid}, respectively, in the Appendix.

\subsection{Molecular emission at the systemic velocity}

\subsubsection{\HCOp\ and \twCO\ emission}
\label{sec:spatial}

As the distance of \snr\ has been determined, we 
study the properties of the neighboring MCs.
Figure~\ref{fig:hcop_co} shows the distribution of \HCOp~\Jotz,\twCO~\Jotz, \twCO~\Jtto, and \HCOp/\twCO~\Jotz\ ratio at around the systemic velocity of the SNR.
Although \HCOp\ and \twCO\ emissions are found in the interior of \snr, they might arise from background or foreground MCs, given the lack of physical evidence for an association with the SNR.
Only the southwestern region has been found to show broadened \HCOp\ emission (see \S~\ref{sec:broad}). 

It is of interest to learn whether SNRs can influence the ISM outside its radio boundary and how the influence is reflected in the chemistry of the MCs.
We found enhanced \HCOp/\twCO~\Jotz\ ratios near the eastern H$_2$ filament, and two clouds in the northeastern and southwestern regions, while the ratio is $\lesssim 10^{-2}$ in the SNR interior.
Hereafter we focus on the $2.12~\um$ H$_2$ filament $0\farcm{5}$ ($\sim 1.2$~pc) away from the eastern radio boundary of \snr\ \citep[see Figure~\ref{fig:4band} and][]{keohane07}.
\footnote{Enhanced \HCOp/\twCO\ intensity ratios are found in two clouds in the northeast and southwest of \snr. 
The two clouds are warmer than 20~K
and have the largest column densities in the FOV.
Our observations cannot tell whether the two clouds are physically associated to the SNR or unassociated star-forming regions, and thus we do not discuss these clouds. 
Moreover, given that the \twCO\ emission is optically thick, a high \HCOp/CO intensity ratio does not necessarily mean a high abundance ratio.}
The near-IR H$_2$ emission requires hot or irradiated molecular gas in a high excitation state, raising questions about the heating mechanism of this filament  \citep[see discussions in ][]{keohane07,lee20}.

To test whether the gas at the eastern $2.12~\um$ H$_2$ filament is shock heated,
we extracted \HCOp\ \Jotz\ spectra at regions D, E, and F.
The line centroid of the $2.12~\um$ H$_2$ emission is at $\VLSR=64.3$--$65.5~\km\ps$ \citep{lee20}.
At similar velocities,
we found \HCOp\ lines with
narrow line widths ($\Delta v=2$--4~\kms), which argues against 
shock perturbation for producing the \HCOp\ emission. 
The narrow line width is not unexpected,
since the SNR's forward shock as traced by the radio emission does not reach the H$_2$ filament and thus should not directly impact the molecular gas. 
Another \HCOp\ peak is found at $\VLSR\sim 58~\km\ps$ to the SNR east
(regions D, E, and F). 
There is no clear evidence to support an association between this velocity and the SNR, given the normal \HCOp/\twCO\ line ratio ($\sim 0.02$) and line widths.

\begin{figure}
	\includegraphics[width=\columnwidth]{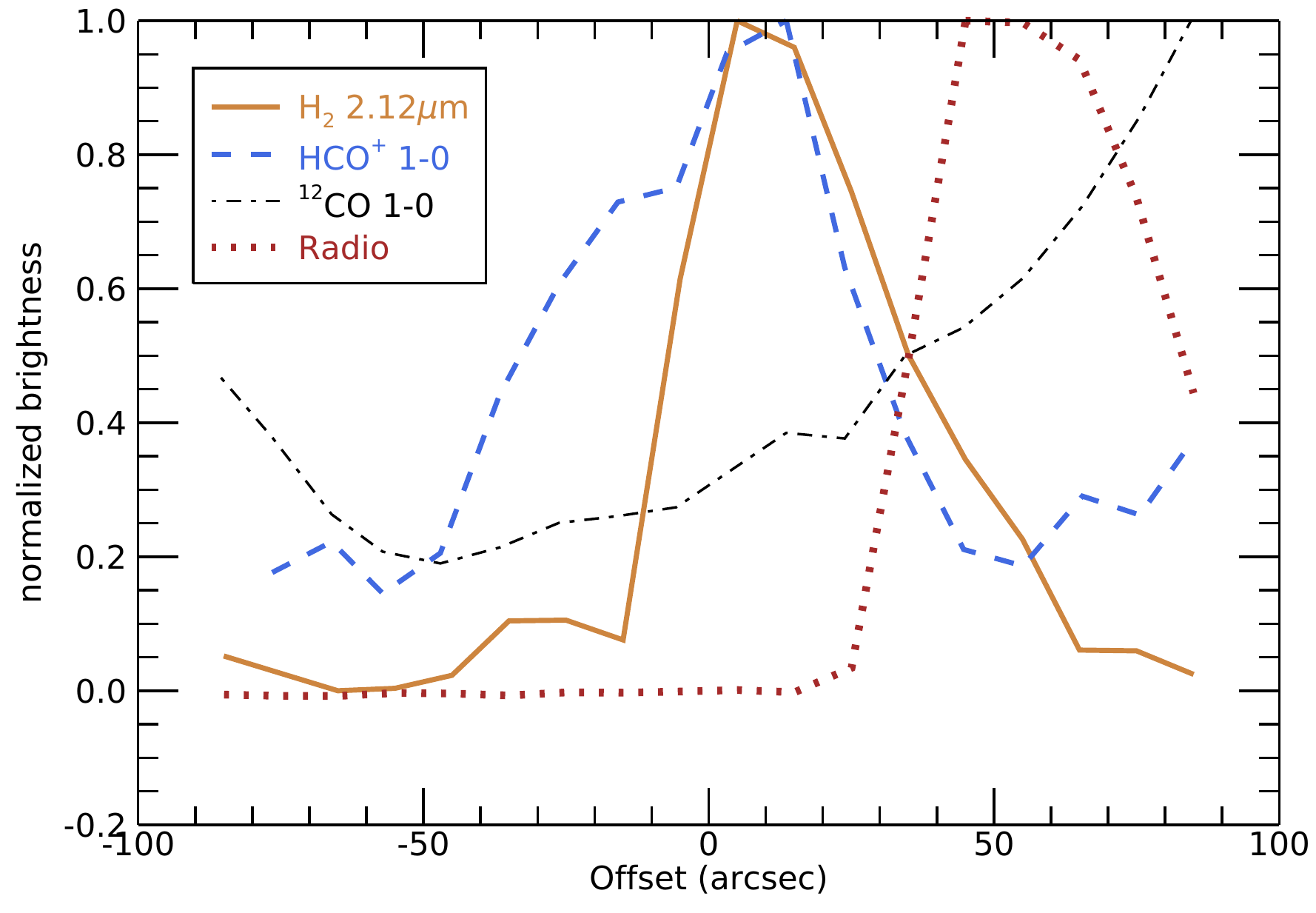}
    \caption{
Brightness distribution of H$_2$ 2.12~$\mu$m, \HCOp, and \twCO, and 327~MHz radio emission cut cross the SNR eastern shell from the east to the west (angle=$30^\circ$; see the region in Figure~\ref{fig:4band}).}
    \label{fig:cutline}
\end{figure}

The eastern \HCOp\ gas with an unusually high \HCOp/\twCO\ ratio extends even further away from the SNR than the H$_2$ filament, although an \HCOp\ knot is seen between the filament and the SNR radio boundary. 
The \twCO\ \Jtto\ map with a better angular resolution also shows a filamentary structure east of the H$_2$ filament.
The layered distributions of different molecular emissions can be more clearly seen in Figure~\ref{fig:cutline}, which is cut along a slice across the eastern SNR shell (see the black dashed box in Figure~\ref{fig:4band}).
Both \HCOp\ and H$_2$ emission peaks lie $\gtrsim 50''$ ($1.9$~pc) away from the radio peak.
The H$_2$ emission is sharply enhanced on the western surface of the \HCOp\ emission, while 
the \twCO~\Jotz\ profile has little resemblance with the H$_2$ profile.
In Section~\ref{sec:irradiated}, we will discuss  how the distribution is explained by the irradiated gas.

\subsubsection{HCN lines and other molecular lines} \label{sec:hcn}

We found HCN~\Jotz\ emission toward \snr, which  
contains a hyperfine triplet due to the nuclear quadrupole moment of N
($F=2$--1, $F=1$--1, $F=0$--1). 
The separation of the hyperfine lines between highest and lowest frequencies is 3.522~MHz, which corresponds to 11.8~\kms.
The brightness pattern among the hyperfine structures is not constant but depends on the excitation condition \citep[e.g.,][]{walmsley82,loughname12}.

Due to the difficulty of separating hyperfine structures, we only examine the HCN properties of the six regions where enhanced \HCOp\ are shown (see regions in Figure~\ref{fig:4band}).
We integrated the HCN~\Jotz\ emission at $\VLSR=40$--150~\kms\ and tabulated the values of the six regions in Table~\ref{tab:lines}.
The flux ratio of \HCOp/HCN is 1.9--2.6 in the shocked gas and $\sim 1$ near the eastern H$_2$ filament.

The enhancement of \HCOp/HCN~\Jotz\ flux at the shocked regions is an intriguing phenomenon, which was also found in SNR IC~443~\citep[with a ratio of $\sim 3$,][]{denoyer81}.
The explanation of the enhancement requires further investigation, which is outside the scope of this paper.

We also detected 
HNC~\Jotz\ at $\sim 64~\km\ps$ at regions A--F.
SiO~\Jtto\ and DCO$^+$~\Jtto\ lines are not detected in our observations.

\section{Discussion} \label{sec:discussion}
\subsection{Large \HCOp/CO ratios in \snr}

\subsubsection{Observed intensity ratios}
We found large \HCOp/CO intensity ratios of 0.2--1 in both the shock-broadened lines in the southwest and
narrow lines outside the SNR eastern boundary (see Figure~\ref{fig:spec}),
while the typical intensity ratios in other molecular components along the line of sight are $\lesssim 10^{-2}$.
This is observational evidence to support that SNRs can influence the chemistry not only of the shocked clouds but also of MCs that are over 1~pc far away.

In the shocked regions, the \HCOp/CO intensity is enhanced at the broadened line wings.
In the velocity range of 75--150~\kms\ (without strong narrow \twCO\ lines), the integrated \HCOp\ intensities are $0.66\pm 0.14$, $1.28\pm 0.11$, and $1.74\pm 0.16~\K \km\ps$, respectively, at regions A, B, and C, while the values for \twCO\ \Jotz\ lines are $0.49\pm 0.33$, $1.84\pm 0.40$, and $1.55\pm 0.31~\K \km\ps$, respectively.
The \twCO\ emission has only a 4--5$\sigma$ detection at regions B and C. 
One cannot claim detection of broad \twCO\ emission in region A as it is below 2$\sigma$, 
We estimate the flux ratio of $I({\rm HCO^+})/I({\rm CO})=1.1\pm 0.4$ at region C and $0.70\pm 0.16 $ at region B.
Given the large velocity gradient, it is reasonable to assume the optically thin case for the \HCOp\ and \twCO\ emission in the broad lines.

Near the eastern H$_2$ filament,
the line ratio of \HCOp/CO 
is large at $\VLSR\sim 64~\km\ps$ compared to those at other velocities.
At region D, the main-beam temperatures of \HCOp\ and \twCO\ are  0.22~K and $\le 1.0$~K, respectively, corresponding to
an \HCOp/\twCO\ intensity ratio $\ge 0.2$. 
Here the \twCO~\Jotz\ main-beam temperature is regarded as an upper limit, as it is in a line valley and the emission is contaminated by the red wing of the unassociated component at $\VLSR\sim 58~\km\ps$.
The \twCO~\Jotz\ emission is probably optically thin, as indicated by a large \twCO~\Jtto/\Jotz\ $\sim 1.5$ \citep[only $\lesssim 1 $ in the optically thick case in the local thermodynamic equilibrium case; see][]{zhou18b}.

The \HCOp/CO intensities obtained 
above are subsequently used to estimate the abundance ratios (see \S~\ref{sec:ratios}), which
will be further compared with those predicted by the thermochemical models of molecular gas in the SNR environment (see \S~\ref{sec:models}).

\subsubsection{Column densities and abundance ratios} \label{sec:ratios}

On the assumption of local thermodynamic equilibrium (LTE) and optically thin emission, the \HCOp\ \Jotz\ line can be used to estimate the column density  $N($\HCOp) \citep{mangum15}:

\begin{multline} 
    N({\rm HCO}^+)\approx 2.46\times 10^{11} \frac{(\Tex +0.78)\exp\left(\frac{h\nu}{k\Tex}\right)}{\exp\left(\frac{h\nu}{k\Tex}\right)-1} \\
    \frac{\int T_{\rm mb} ({\rm HCO^+}) dv (\km\ps)}{J_\nu (\Tex)-J_\nu (T_{\rm bg})} \cm^{-2}
\end{multline}

\begin{figure*}
\centering
	\includegraphics[width=0.45\textwidth]{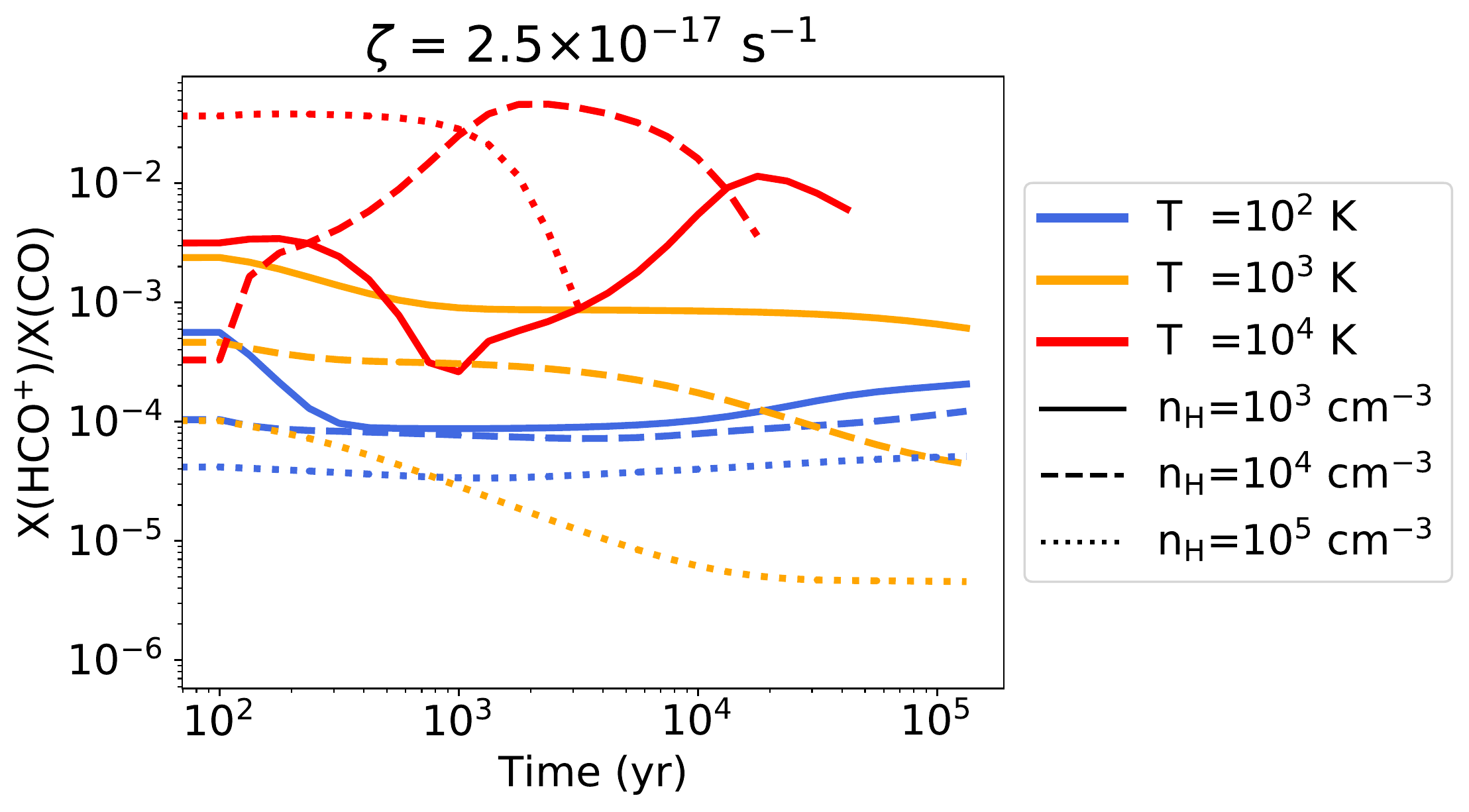}
	\includegraphics[width=0.45\textwidth]{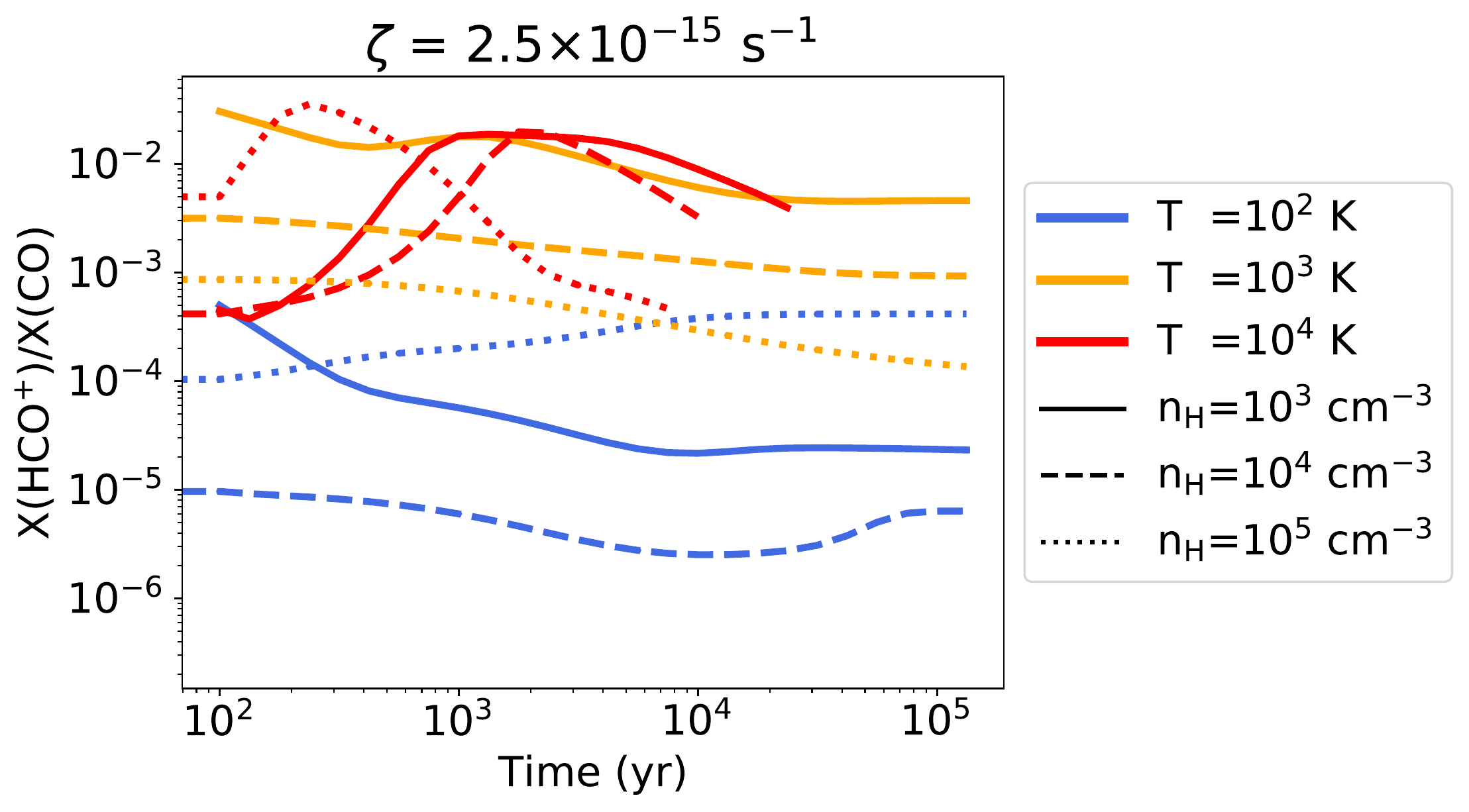}
	\includegraphics[width=0.45\textwidth]{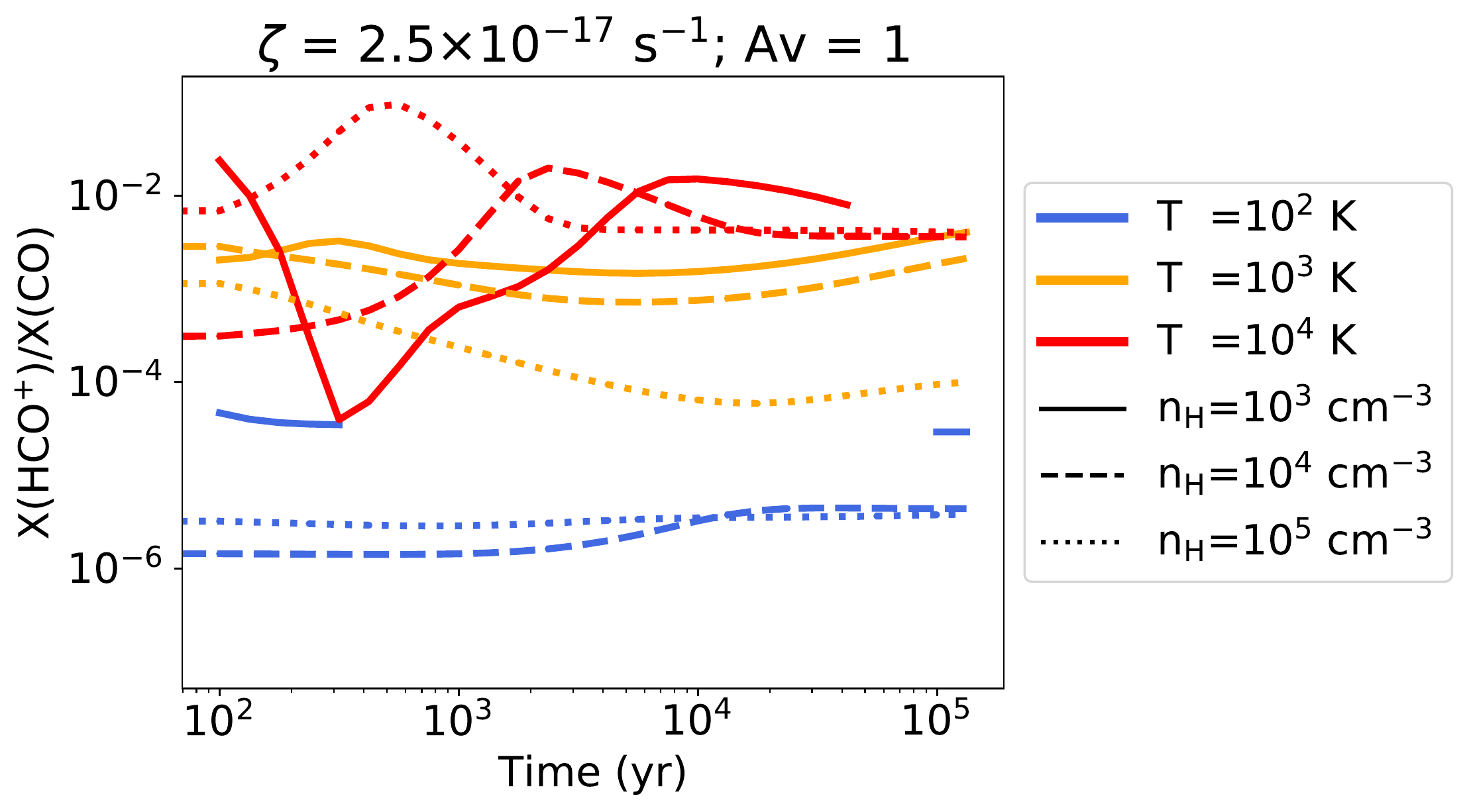}
	\includegraphics[width=0.45\textwidth]{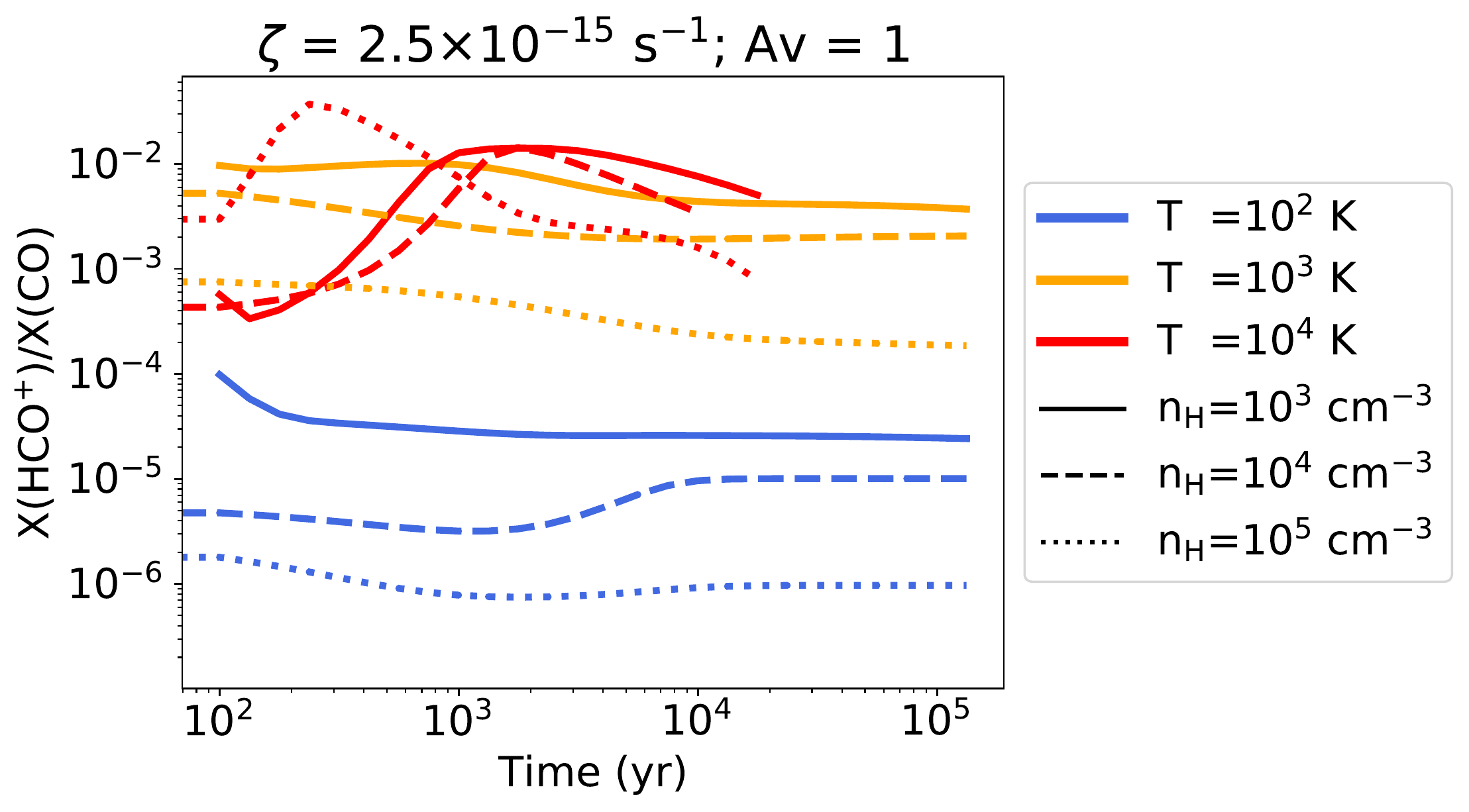}
	
    \caption{
    The chemical evolution in the shocked molecular region.
    $\zeta$ is the CR ionization rate per H$_2$. 
    The number density of hydrogen $\nH$ is contributed by atoms, ions, and molecules. 
    $A_{\rm v}$ is the extinction magnitude of the standard Draine UV photon field to which the MC is exposed.
    In the upper row, the effect of the UV photon field is assumed to be extremely small. We discard the ratio values when either the number density of \HCOp\ or CO is lower than $10^{-8}\cm^{-3}$.
	}
    \label{fig:chem}
\end{figure*}

where $h$ and $k$ are Planck constant and Boltzmann constant, respectively, $\nu$ is the frequency of the \Jotz\ transition, $J(T)=\frac{h\nu}{\exp(h\nu/kT)-1}$, and the Cosmic Microwave Background temperature $T_{\rm bg}=2.73~\K$. For a kinetic/excitation temperature $\Tex\gg T_{\rm bg}$, the above equation is simplified as

\begin{multline} \label{Eq:hcop}
N({\rm HCO^+})\sim 5.7 \times 10^{10} \Tex \exp\left(\frac{4.28}{\Tex}\right) \\
\int \Tmb ({\rm HCO^+}) dv (\km\ps)
\cm^{-2}
\end{multline} 

Similarly, we derive the \twCO\ column density using the optically thin \twCO\ \Jotz\ emission for $\Tex \gg T_{\rm bg}$:

\begin{multline} \label{Eq:co}
    N({\rm CO})\sim 4.32 \times 10^{13} \Tex \exp\left(\frac{5.53}{\Tex}\right) \\
    \int \Tmb ({\rm CO})dv (\km\ps) \cm^{-2}
\end{multline} 

From equations~(\ref{Eq:hcop}) and (\ref{Eq:co}), it is
convenient to obtain $\NHCOp/N({\rm CO}) \sim 1.3\times 10^{-3}  I({\rm HCO^+})/I({\rm CO}) \sim 1.3\times  10^{-3}  \Tmb ({\rm HCO^+})/\Tmb({\rm CO})$ for a gas excitation temperature $\Tex\gtrsim 20~\K$.

We estimate the abundance ratio $X({\rm HCO^+})/X({\rm CO})=N({\rm HCO^+})/N({\rm CO})=3\times 10^{-4}$--$2\times 10^{-3}$ 
at the regions with enhanced \HCOp/CO intensity ratios.
The abundance ratios are significantly larger than those found in cold MCs ($10^{-6}$--$10^{-4}$), where the fractional abundance of CO relative to H$_2$ \citep[$X({\rm CO})\sim 10^{-4},$][]{dickman78,lucas98} is a few orders of magnitude larger than that of \HCOp\ 
\citep[$X({\rm HCO^+})\sim 10^{-10}$--$10^{-8}$ varies with excitation conditions;][]{miettinen14,fuente19}.

We caution that the \HCOp/CO abundance ratios are obtained by assuming an LTE condition, considering we only have one \HCOp\ transition.
Further observations with mid-J and high-J transitions of \HCOp\ and \twCO\ are needed for a precise diagnosis of the excitation conditions.

\subsection{Model the chemical and physical properties of a shocked MC exposed to CRs} \label{sec:models}

The SNR shock can heat and ionize the gas 
and change the abundance of molecular species.
The GeV--TeV observations of \snr\ have proved that the SNR is accelerating CR protons 
\citep{hess18},  
which are  known as important sources regulating molecular chemistry. 
We used two codes to compute the \HCOp/CO abundance ratios of a shocked MC exposed to CRs.

\subsubsection{Constant-temperature chemical model}

We first used a chemical network with molecules containing H, C, O, and N \citep{Bovino16,hily18, legal14} to simulate the chemical evolution in a shock-heated cloud by employing the astrochemistry package KROME \citep{Grassi2014}. The initial condition is assumed to be in equilibrium at a temperature $T=15 \K$ for various densities.
We then calculate the evolution of every species after an abrupt increase of the temperature to 
about $10^2$ K, $10^3$ K, and $10^4$ K (corresponding to a shock velocity of about $1.4\km\ps$, $5 \km\ps$, and $15~\km\ps$, respectively; see Figure~\ref{fig:chem}). In the upper row of Figure~\ref{fig:chem}, we ignored the effect of photoionization. In the lower panels, we used an extinction level, $A_V=1$, 
with a standard Drain UV field \citep{draine1978}. 
The exact radiation field and extinction near \snr\ have been thus far unclear (see \S~\ref{sec:irradiated}).
By setting a moderate extinction level $A_V$ to 1, we assume the column density for shielding the radiation field is about $1.9\times10^{21}~{\rm cm}^{-2}$ \citep{Bohlin78} to show the effect of a moderate UV photoionization condition. 
We set a lower limit of the molecular number density, $10^{-8}\cm^{-3}$, to make sure that the molecular emission are observable and prevent a small denominator.

According to our network, 
at the temperature of $10^3\K$, the main channels to produce \HCOp\ are 
CO + H$_3^+$ $\rightarrow$ H$_2$ + \HCOp and HOC$^+$ + H$_2$ $\rightarrow$ \HCOp\  + H$_2$. 
The former reaction converts some CO molecules to \HCOp, raising the abundance ratio between \HCOp\ and CO. 
Figure~\ref{fig:chem} reveals
a higher \HCOp/CO ratio for enhanced CRs,
which can increase the number of ions in the MCs.
When the temperature is raised to $10^4\K$, both CO and \HCOp\ molecules can be easily dissociated, but
the CR ionization reaction chain (such as the reaction HCO $\xrightarrow{\rm CR}$ \HCOp\ + $e^-$ and CO + H$_3^+$ $\rightarrow$ \HCOp +  H$_2$)
keeps the \HCOp\ dropping slower than CO, resulting in an \HCOp/CO abundance ratio $> 10^{-3}$. 
In a slow shock with a lower temperature of $10^2$ K, 
the \HCOp/CO abundance ratios are below $10^{-3}$ for various densities. Although the CR reaction chain still produces \HCOp,
the recombining reaction of \HCOp\ (e.g., \HCOp $+ e^- \rightarrow $ CO $+$ H) 
and the reactions between \HCOp\ and neutral atoms or molecules (e.g., \HCOp $+$ C $\rightarrow$ CO $+$ CH$^+$) are relatively fast in this low temperature, resulting in a low \HCOp/CO abundance ratio. 

We note that this constant-temperature model is a simplified model that does not include cooling/heating mechanisms or the the magnetic fields, and the chemical network only contains up to four atom molecules. 
Nevertheless, this model shows that \HCOp/CO abundances are highly sensitive to gas physical condition and CR ionization rate. This model is also useful to learn the main channels to form \HCOp\ in a given condition.

\subsubsection{MHD shock model} \label{sec:mhd}

We further simulated the \HCOp/CO ratio in a magnetohydrodynamic (MHD) shock 
using the Paris--Durham shock code \citep{godard19}, which takes the influence of magnetic fields on shocks into account.
The MHD shock code calculates the physical and chemical evolution of the gas, by including over 125 species linked by more than 1000 reactions and with cooling and heating processes considered. 
We considered a plane-parallel shock wave 
propagating in a molecular and dusty homogeneous medium with 
a magnetic field strength of $B=\left(\frac{\nH}{\cm^{-3}}\right)^{1/2} \mu $G, where 
$\nH$ is the preshock gas density of the three-phase gas $n_{\rm H}=n({\rm H})+2n({\rm H_2})+n({\rm H^+})$.

Firstly, we calculated the thermochemical conditions of the preshock gas, adopting the standard initial distribution of elemental fractional abundances among gas-phase species, PAHs, icy grain mantles, and grain cores, as adopted by \cite{flower03}.
We allow the gas to evolve in a quiet and dark condition with  given densities and the Galactic level CR ionization rate $\zeta=2.5\times 10^{-17}~\ps$ per H$_2$ \citep[a few $10^{-17}$~s$^{-1}$ in dense MCs;][]{vandertak00}. The final static thermochemical properties are
used as input parameters for the preshock gas.

\begin{figure}
\centering
	\includegraphics[width=0.5\textwidth]{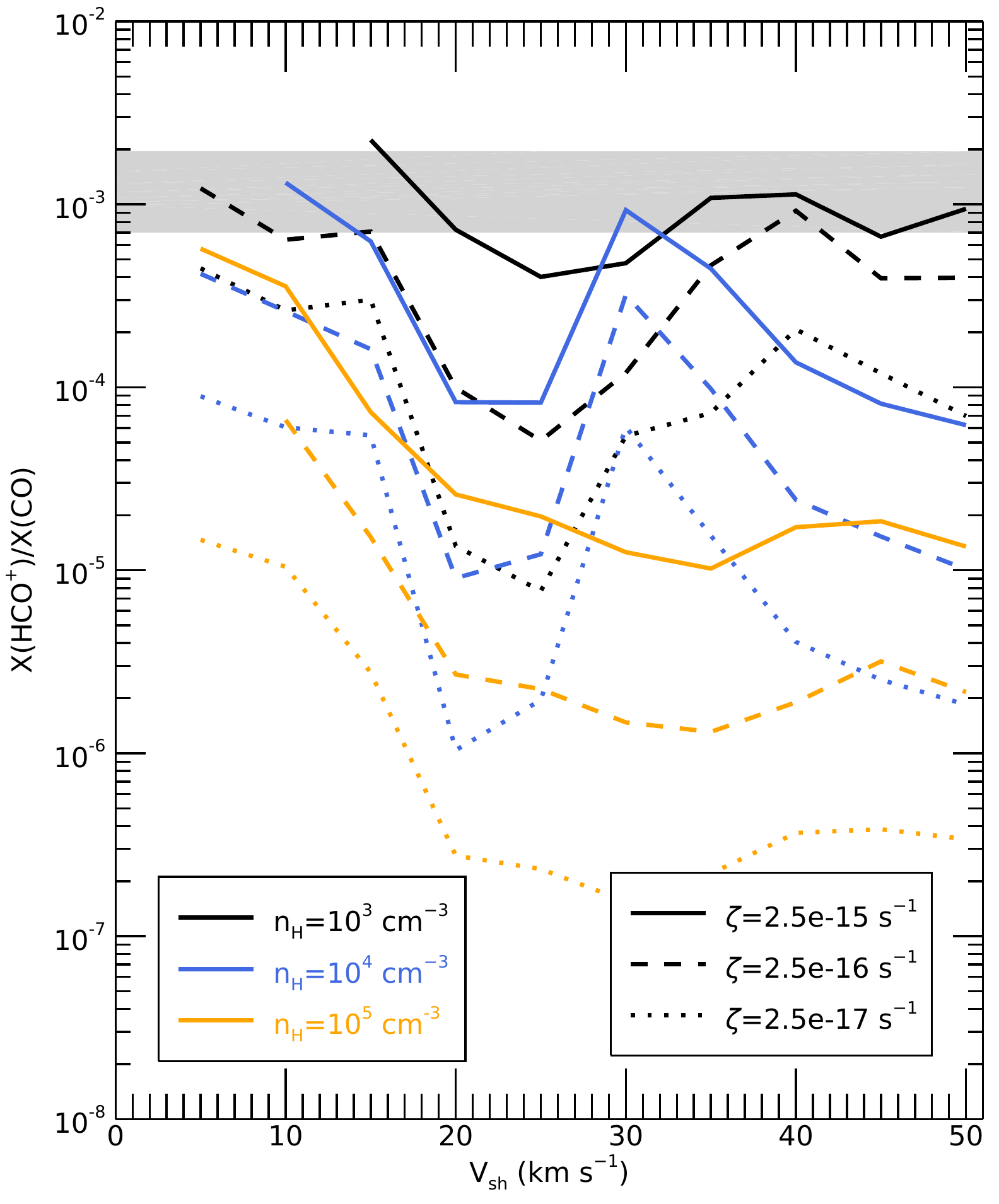}
    \caption{
    The abundance ratio of \HCOp/CO as a function of entrance shock velocity, preshock gas density n$_{\rm H}$, and CR ionization rate per H$_2$ $\zeta$ in the MHD shock models. The shock age is taken as 1~kyr. The observed values in regions B and C are denoted by the gray area.
	}
    \label{fig:mhd_shock}
\end{figure}

Subsequently, we run the MHD shock models, which
cover entrance shock velocities of 5--50~\kms\ (a step of $5~\km\ps$), and preshock densities of $10^3$, $10^4$, and $10^5~\cm^{-3}$.
The Galactic level ($\zeta=2.5\times 10^{-17}~\ps$) and higher levels ($\zeta=2.5\times 10^{-16}~\ps$, $\zeta=2.5\times 10^{-15}~\ps$) of CR ionization rates are considered to evaluate the CRs' influence. 
We adopted a shock timescale of 1~kyr in MCs. 
With an age of  5--6~kyr \citep[][]{zhou18a,sun20}, \snr\ 
should not have spent its whole lifetime interacting with MCs.
Moreover, the molecular shock timescale is consistent the recombination age of the cooler X-ray plasma in the southwestern region of \snr\ 
\citep[$\sim 1$--2~kyr,][]{zhou18a,hollandashford20}, 
where the thermal conduction with the cold and dense gas likely caused a rapid cooling of the X-ray-emitting plasma.

The simulations show that the shock types 
at $V_{\rm sh} < 20~\km\ps$
are nonstationary ``continuous'' shocks (young C-type shock), which are composed of a magnetic precursor and J-type tail \citep[see][for the temporal evolution of young C-type shocks]{lesaffre04b}.
The shocks at $V_{\rm sh} \gtrsim 20~\km\ps$
become ``jump'' shocks (J-type), where the ions
and neutrals are coupled.

\begin{figure*}
\centering
	\includegraphics[width=0.48\textwidth]{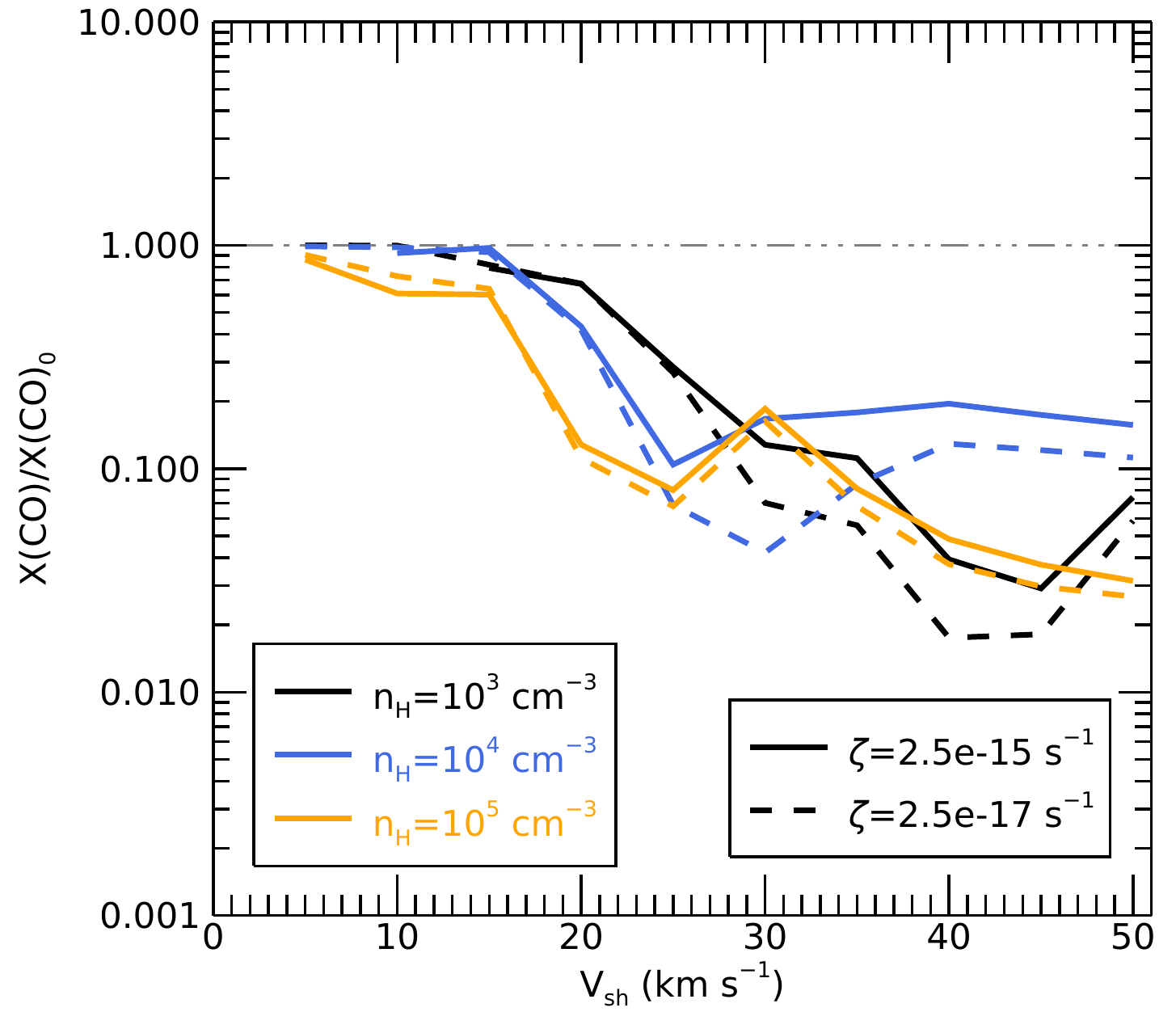}
	\includegraphics[width=0.48\textwidth]{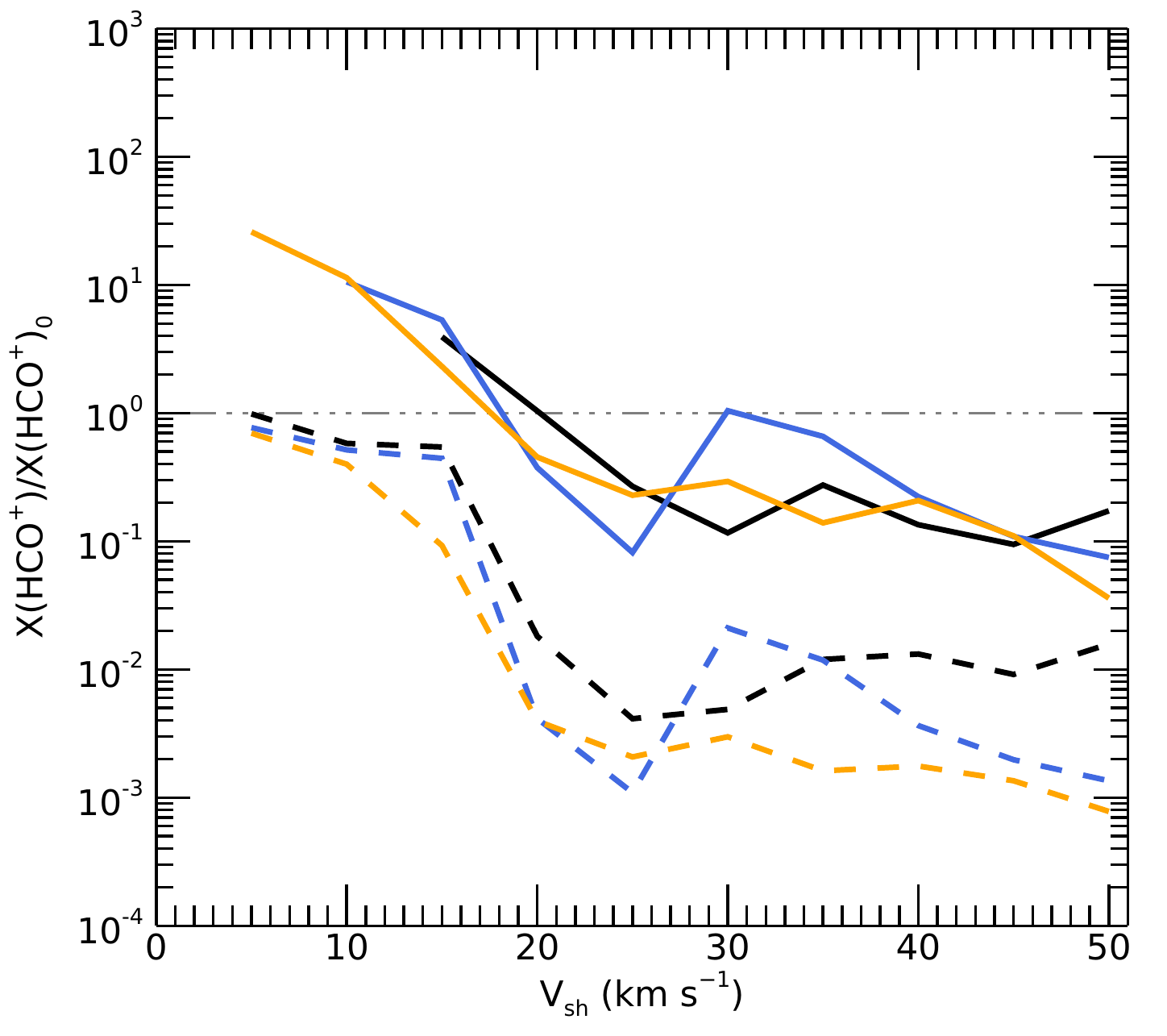}
    \caption{
    The abundance ratio of \twCO\ (left) and \HCOp\ (right) relative to those in the preshock gas for a shock age of 1~kyr.
	}
    \label{fig:abun}
\end{figure*}

Figure~\ref{fig:mhd_shock} shows the modeled abundance \HCOp/CO ratios with a shock age of 1~kyr,
as a function of shock velocity, preshock gas density, and cosmic-ray ionization rate.
The abundance ratio
is obtained from the integrated \HCOp\ and \twCO\ column densities in the shocked layer.
 $X[{\rm HCO^+}]/X[{\rm CO}]$ tends to decrease with increasing preshock gas density and shock velocity, while the
 enhanced CR ionization rate can  significantly increase the ratio.

A high CR ionization and a moderate density are needed to explain the observed large $X({\rm HCO^+})/X({\rm CO})$, while the Galactic level CR  (dotted lines) and very dense preshock gas (orange color) cases predict a too-low value.
It is reasonable to expect two orders of magnitude higher CR ionization rate $\zeta$ near \snr, as found in other middle-age SNRs such as IC443 \citep[$\zeta=2\times 10^{-15} \ps$;][]{indriolo10} and W28 \citep{vaupre14}. 
With $\zeta=2.5\times 10^{-15}~\ps$,  the high $X({\rm HCO^+})/X({\rm CO})$ can be reached in a wide range of shock velocities (see Figure~\ref{fig:mhd_shock}).
At $V_{\rm sh} \ge 25~\km\ps$, $X({\rm HCO^+})/X({\rm CO})$ is less sensitive to the shock velocity  but relies more on the CR ionization rate and density.

The broad \HCOp\ lines ($\Delta v=50$--75~\kms) found in \snr\ hint at a fast shock in the MC. The broad-line profile can be explained by either a single cloud with a large velocity dispersion or multiple shocked  structures accelerated to different LSR velocities. In either case, a large shock velocity is needed to explain the large line centroid shifts (from $\VLSR\sim 60~\km\ps$ of the quiescent gas to $\VLSR\sim 100~\km\ps$ for gas in region C).

We suggest that the high \HCOp/CO ratio in the broad-line regions can result from a CR-induced chemistry in shocked MCs with
a preshock density $\nH\sim 10^3$--$10^4~\cm^{-3}$. 
An enhancement of the HCO$^+$/CO ratio induced by CRs has also been found in a low-to-intermediate density cloud, IC~348 (Luo et al.\ in preparation).
We do not claim an enhancement of the \HCOp\ abundance, but only an enhancement of the \HCOp/CO abundance ratio.
Our MHD shock modeling shows that
the \twCO\ molecules tend to be dissociated with enhanced shock velocities. \twCO\ abundance can be 1--2 orders of magnitude lower than the level in the
preshock gas with $V_{\rm sh}\gtrsim 20~\km\ps$ (see Figure~\ref{fig:abun}).
The \HCOp\ abundance also presents a negative trend to the shock velocity, but the high-level CR ionization strongly regulates the \HCOp\ chemistry and maintains it above 0.1 of the preshock level.
When shock velocity is low ($V_{\rm sh} \lesssim 15~\km\ps$ ) and does not effectively destroy \HCOp\ molecules, the enhanced CRs can indeed cause an \HCOp\ abundance enhancement.
Nevertheless, the low shock velocities are unlikely to explain the  broad \HCOp\ profiles and the weak CO emission.
Therefore, the observed high ratio of \HCOp/CO  does not mean an enhancement of \HCOp, but
reflects the more efficient destruction of CO molecules than that of the \HCOp\ in the faster shock.

There are a few caveats in our MHD modeling and comparison.
We have not evaluated the influence of external radiation on the \HCOp/CO abundance ratio, but only show that the enhanced CRs can already explain the observation. The UV spectrum of \snr\ is unclear, while the X-ray-induced physics/chemistry is not considered in the MHD models. 
Moreover,  the observed $X({\rm HCO^+})/X({\rm CO})$ for comparison is calculated under a simplified LTE assumption.

\subsection{Irradiated gas in the SNR east}

An unusually high \HCOp/CO line ratio is also found over 1~pc away from the SNR eastern boundary (see Figure~\ref{fig:spec}).
Here we discuss three types of precursors that might cause peculiar chemical properties to the MCs ahead of the shock (near the eastern H$_2$ filament; see Figure~\ref{fig:4band}): 1) CR precursor, 2) magnetic precursor, and 3) radiation precursor.
We do not consider the SNR blast wave as the direct source, as it is too far from these regions to cause the \HCOp/CO enhancement. 
The narrow \HCOp\ line widths do not suggest the shock perturbation, either.

\subsubsection{CR precursor}

CRs can diffuse away from the SNR boundary. The diffusion length of CRs is expressed as \citep[see][and references therein]{vink12}:

\begin{multline}
l_{\rm diff}=1.0\E{-2} \pc 
~\eta^{1/2}\left(\frac{E}{{\rm 1~GeV}}\right)^{1/2}\left(\frac{B}{10 \mu {\rm G}}\right)^{-1/2} \\
\left(\frac{t}{5~{\rm kyr}}\right)^{1/2}
\end{multline}

where $\eta$ accounts for the deviation from Bohm diffusion \citep[$\lesssim 10$ in young SNRs,][]{vink12}, $E$ is the CR particle energy, $B$ is the magnetic field and $t$ is the acceleration age, equal to the SNR age (5--6~kyr).

Low-energy CRs ($<1$~GeV) cannot diffuse over $ 1$~pc away from the shock, unless $\eta$ is unusually large. 
High-energy CRs with $E\ge 1$~TeV could reach this distance, but their energy losses via pion production 
are a few orders of magnitude larger than
ionization losses, meaning inefficient in MC heating \citep{mannheim94}.
Assuming a hadronic origin of the $\gamma$-ray emission from \snr,
$\gamma$-rays carry about at least 1/3 of the pion production losses, and 
the spectrum  peaks at about 1/10 of the energy of the primary CRs \citep{hinton09}.
The total $\gamma$-ray flux of \snr\ above 0.1~TeV is $F(>0.1~{\rm TeV})\sim 4\times 10^{-12}~\erg~\cm^{-2}~\ps$ \citep{hess18}, 
only comparable to the flux of the 2.12~$\mu$m H$_2$ line east to the SNR \citep[$F(2.12\mu {\rm m~H_2})\sim 5\times 10^{-12}~\erg~\cm^{-2}~\ps$;][]{lee19}.
The comparison suggests that the ionization losses from TeV CRs of \snr\ are insufficient to power the observed molecular emissions.
Therefore, CRs are not the favored heating source for the clouds east to \snr. 

\subsubsection{Magnetic precursor}

A magnetic precursor can appear in weakly ionized MCs, which coexist two distinct fluids in the shocks: a neutral fluid and an ionized fluid.
If the magnetosonic speed exceeds the shock speed ($V_m > V_{\rm sh}$), compressive waves can be generated upstream ahead of the shock.
The extent of the magnetic precursor for a stationary C-type shock is expressed
as \citep{draine11}:

\begin{multline}
\label{eq:lmag}
    l_{\rm mag}\approx 3.4\times 10^{15}~\cm \left(\frac{B}{10~\mu {\rm G}}\right)^2 \left(\frac{V_{\rm sh}}{10 \km\ps}\right)^{-1} \\
    \left(\frac{\nH}{10^2 \cm^{-3}}\right)^{-2} \left(\frac{x_i}{10^{-4}}\right)^{-1}
\end{multline}

where the fractional ionization $x_i$ is  $\sim 10^{-4}$ in the diffuse MCs. 
For a dark cloud with a density of $10^2$--$10^5~\cm^{-3}$, there is a simple approximation $x_i\approx 10^{-5} (\nH/\cm^{-3})^{-1/2}$ \citep{draine11}. 
The ionization fraction is raised
in clouds exposed to enhanced UV/X-ray radiation or CRs, which could happen near \snr.
Assuming the magnetic energy conservation,
the magnetic field in the cloud is estimated as $B=b_0 (\nH/\cm^{-3})^{1/2} \mu {\rm G}$.
Equation~\ref{eq:lmag} is then rewritten
as $l_{\rm mag}\sim 3.4\times 10^{19}~\cm~b_0^2[(\nH/
\cm^{-3}) (V_{\rm sh}/ \km\ps)^2]^{-1/2}$.
A previous X-ray study found a blast wave velocity of $V_{\rm bw}\sim 480~\km\ps$ in \snr, with 
a preshock intercloud density of $n_0\sim 10~\cm^{-3}$ \citep{zhou18a}.
Assuming a pressure equilibrium between the blast wave and cloud shock $\nH V_{\rm sh}^2\sim n_0 V_{\rm bw}^2$ \citep{mckee75},
the expression of the magnetic precursor length for \snr\ can be simplified
as:
$l_{\rm mag}\sim 2\times 10^{16} b_0^2~\cm$, where
$b_0$ is in the level of $\sim 1$ in typical MCs \citep{crutcher10} but could vary with specific environments.
It was proposed that the magnetic precursor causes the high-excitation H$_2$ emission outside the SNR boundary \citep{keohane07}. 
However, our calculation suggests that the magnetic precursor length of \snr\ is too small to excite the H$_2$ and \HCOp\ emission 1~pc away from the SNR boundary, while \snr\ is unlikely to be located in a strongly magnetized medium \citep{brogan01}.

Our MHD models reveal that shock types in the southwestern MCs are not stationary C-type shocks, but
are likely J-type with shock velocities $\gtrsim 20~\km\ps$ (see Section~\ref{sec:mhd}).
As the ions and neutrals are coupled in the J-type shocks, the magnetic precursor can be ignored.

\subsubsection{Radiation precursor} \label{sec:irradiated}

Unlike the CR and magnetic precursors that have short-length scales, radiation can freely escape the SNR until the photons are absorbed/scattered by the ISM.
\snr\ is the most luminous Galactic SNR in the X-ray band \citep{immler05} with an intrinsic luminosity of $L_{\rm X}\sim 2$--$9\times 10^{38} \du^2~\erg\ps$  in 0.3--10 keV band \citep[the upper limited is recalculated using the XMM-Newton data and model in][]{sun20}, 
where the uncertainty mainly comes from different absorption models.
Its soft X-ray emission is so heavily absorbed 
that the observed 0.3--10~keV X-ray luminosity is only
$5\E{35}~\erg\ps$.
This means that most of the soft X-ray energy is transferred to the environment
of \snr\ and the ISM along the line of sight.

Previous X-ray spectroscopy
reveals that the soft X-ray emission below 1~keV is contributed by hot plasmas with
a temperature of $kT\sim 0.2$--0.3~keV \citep[$\sim 3\E{6}~\K$][]{zhou18a,sun20,hollandashford20}.
If the soft X-ray emission is characterized by an $apec$ model with a temperature of 0.177~keV and a normalization of $86.7~\cm^{-5}$(see the best-fit $apec+2*vvrnei$ model in \cite{sun20}), we can roughly estimate a UV luminosity of $L_{\rm UV}\sim 6\times 10^{39}~\erg\ps$ at 6--300~eV and $7\times 10^{37}~\erg\ps$  at 6--13.6~eV.
The values are likely lower limits, as there could be undetected gas components cooler than $\sim 0.2$~keV.
Therefore, \snr\ is also a luminous UV source.

Given the high X-ray and UV luminosity of \snr, we suggest radiation precursor be a plausible source for the hot MCs to the SNR east. 
As the UV emission is highly absorbed, we do not clearly know the  UV spectrum of \snr\ and thus do not evaluate its influence on the MCs outside \snr. 
However, 
the strong UV emission can explain the existence of strong 8~$\um$ IR emission to the SNR east (see Figure~\ref{fig:4band}), which traces
either the polycyclic aromatic hydrocarbons (PAHs) in strong radiation fields or hot dust grains \citep{draine11}.
There is a good correlation between the near-IR H$_2$ emission and 8~$\mu$m emission, but the latter emission is even more extended (up to $1\farcm{7}$ or $4$~pc) from the radio boundary. 

The large X-ray luminosity of \snr\  makes it a potential laboratory to study MCs in XDRs.
It has been proposed that the intense X-ray emission 
from SNRs could transfer a fraction of their energy to MCs and give rise to 
strong vibration-rotational H$_2$ \citep{draine90,draine91}.
Indeed, \snr\ is the most luminous SNRs in both the X-ray band and near-IR H$_2$ emission based on our current knowledge \citep{lee19}.
According to XDR models by \cite{maloney96},
the H$_2$ emission in XDRs has 
a relatively low ratio H$_2$ 2-1 $S(1)/$ 1-0 $S(1)$ $\sim 0.1$ ($\lesssim 0.3$), similar to the values in shock excited gas, 
while the mean ratio for UV fluorescent excitation of H$_2$ is $\sim 0.5$--0.6 \citep{black87}.
\cite{lee20} observed a ratio of 0.07--0.10 in
the eastern H$_2$ filament and proposed that the gas could be either shocked or X-ray-irradiated gas.
Here we prefer the latter scenario, as the former case is less likely.

We found that the \HCOp\ emission is generally in the outer layer of the H$_2$ filament to the SNR east (see Figure~\ref{fig:cutline}).
This distribution can also be explained with XDR models, which predict that H$_2$ emission arises from the surface layer of the irradiated MCs \citep{maloney96}.
The layered \HCOp\ and H$_2$ distribution likely reflects a radiation attenuation into the MCs.
The XDR models by \cite{meijerink05} indeed provide multiple solutions for
producing the observed high \HCOp/CO intensity ratio ($\ge 0.2$ to the SNR east). 
However, the X-ray spectral shape ($\exp (-E/1 {\rm keV})$) 
and flux range (1.6, 16, and $160~\erg\cm^{-2}\ps$) in 
the available XDR models do not cover that in \snr\ \citep[see the X-ray spectral models in, e.g.,][]{sun20},
while the observed molecular transitions are too few to constrain the gas excitation condition. A reliable comparison can be provided in the future using new molecular observations and XDR models with flexible X-ray spectral parameters.

\subsection{Remarks about \HCOp/CO ratios near SNRs}

In \snr, optically thin \HCOp\ emission and \HCOp/CO ratio are strongly enhanced in the shocked regions and thus appear to be good tracers of SNR shock under these particular conditions. 
We also note that high \HCOp/CO intensity ratios of $\sim 1$ are found in the shocked gas in IC~443 \citep{dickinson80,dickman92}.
It is of interest to explore the \HCOp/CO ratios in more SNRs that are impacting MCs,
to test whether the high \HCOp/CO ratio can be a good tracer of SNR shock in various physical conditions.
If it is the case, \HCOp\ lines will 
be a useful tool to identify new SNR--MC associations on the Galactic plane.

Broadened \twCO\ emission has been widely used to
probe shocked MCs
and many known SNR--MC interaction has been identified in this way \citep{jiang10}.
However, \twCO\ emission is not a shock tracer, but is usually used as a molecular gas tracer since CO is an abundant molecule in the interstellar MCs and easy to be excited.
In the $\VLSR$ range of 0--80~\kms, the \twCO\ \Jotz\ and \Jtto\ emissions toward \snr\ are subject to the severe line crowding, because multiple quiescent MCs lie along the line of sight. This causes problems to discern any broad profiles.
Although the line of sight contamination should be free 
at $\VLSR> 80~\km\ps$ (see Figure~\ref{fig:spec}), we still
do not detect strong broad \twCO\ emission across the SNR, probably because of the efficient destruction of \twCO\ molecules in the shocks (see Figure~\ref{fig:abun}).

Finally, we observed high \HCOp/CO intensity ratios outside \snr, which
is likely due to the strong X-ray/UV radiation.
It is also of interest to test whether such high ratios can be found outside other SNRs (or other X-ray/UV luminous sources) in the molecular environment.
Nevertheless, the ratio should be the subject to multiple physical parameters (such as temperature, density, and optical depth), in addition to radiation fields.
Further modelings of the MC chemistry in strong radiation are needed for comparison.

\section{Conclusion} \label{sec:conclusion}
We have performed a molecular line survey toward SNR \snr\ using the IRAM 30~m telescope.
In this paper, we focus on \HCOp\ and \twCO\ emissions, 
to probe the influence of SNR on the MCs through shocks, CRs, and radiation.
Our main conclusions are as follows:

\begin{enumerate}
    \item \snr\ is interacting with  MCs in its southwest, where we found very broad \HCOp\ emissions
    with widths of 48--75~\kms.
    The observed flux ratios of \HCOp/CO in 
    two shocked regions are $1.1\pm 0.4$ and $0.70\pm 0.16$, which are significantly larger than that in typical MCs.
    In the LTE condition and optically thin case, the abundance ratios are estimated as 
    $X({\rm HCO^+})/X({\rm CO})=7\times 10^{-4}$--$2\times 10^{-3}$. We have also found an enhanced flux ratio of \HCOp/HCN of 1.9--2.6 in the shocked gas, larger than the ratio of $\sim 1$ near the eastern H$_2$ filament.
    
   \item By comparing with the MHD shock models, 
   we suggest that the high \HCOp/CO ratio in the broad-line regions can result from a CR-induced chemistry in shocked MCs, where the CR ionization rate is enhanced to 1--2 orders of magnitude larger than the Galactic level.
   The observed high \HCOp/CO ratio in the broad lines does not mean an enhancement of \HCOp\ but reflects a stronger decline of the CO abundance than that of \HCOp.

   \item Our molecular study supports the systemic velocity of \snr\ is $\VLSR\sim 61$--$65~\km\ps$ and the distance is $7.9\pm 0.6$~kpc, consistent with that obtained using the near-IR H$_2$ emission \citep{lee20}. Therefore, \snr\ does not have a physical relation to the star-forming region W49A. 
   
    \item 
    SNR \snr\ also influences the molecular chemistry far ahead of its shocks.
    We have found an unusually high \HCOp/CO intensity ratio ($\ge 0.2$) over 1~pc east of the SNR's radio boundary, where the \twCO\ emission is optically thin.
    The \HCOp\ lines outside the SNR are narrow ($dV=2$--4~\kms), disfavoring an SNR shock perturbation.
    The characteristic lengths of
    low-energy CRs and magnetic precursors of \snr\ are much smaller than 1~pc, and thus are unlikely to cause the enhanced ratio while the radiation precursor (X-ray or UV) is capable to change the molecular chemistry outside the SNR. 
    
    \item   
    As the most luminous Galactic SNR in the X-ray band,
    \snr\ is a potential laboratory to study the XDRs. 
    The XDR models can explain the low 
    H$_2$ 2-1 $S(1)/$ 1-0 $S(1)$ $\sim 0.1$ ratio in the H$_2$ filament east to the SNR \citep{lee20}, while PDR models predict larger values.
    Moreover, we found that the \HCOp\ structure is east of the H$_2$ filament. 
    This layered distribution can be explained under XDR models, which predict that H$_2$ emission arises from the surface layer of the irradiated MCs. 
    The layered \HCOp\ and H$_2$ distribution likely reflect a radiation attenuation into the MCs.
    Although some XDR models can explain the observed \HCOp/CO ratio, we caution that
    future models and observations are needed for a reliable comparison.

    \item  The \HCOp/CO ratio is a potentially useful tool to study
    the influence of SNR shocks, CRs and radiation on the MC chemistry.
    The low-J \twCO\ emission alone does not seem to be a good probe for shock conditions near \snr\ on the Galactic plane, as it is weak in  the shocked gas, but too strong in unassociated clouds along the line of sight.

\end{enumerate}

\begin{acknowledgements}
The authors are thankful to the anonymous referee for constructive suggestions. We also thank Xiao Zhang for the helpful discussion about the $\gamma$-ray emission of \snr.
This work is based on observations carried out under project numbers 167-18 and 024-20 with the IRAM 30m telescope. IRAM is supported by INSU/CNRS (France), MPG (Germany) and IGN (Spain).
Shock models published in this paper have been produced with the Paris--Durham shock code
\citep[http://ism.obspm.fr]{flower03,lesaffre13,godard19}.
P.Z.\ acknowledges the support from 
NSFC grant No.\ 11590781, 
Nederlandse Organisatie voor Wetenschappelijk Onderzoek (NWO) Veni Fellowship, grant no. 639.041.647, and the Nederlandse Onderzoekschool Voor Astronomie (NOVA).
M.A. thanks the NWO for support via the Talent Programme Veni grant.
The work of B.-C.K.\ is supported by Basic Science Research Program through the National Research Foundation of Korea funded by the Ministry of Science, ICT and Future Planning (2020R1A2B5B01001994).
Z.-Y.Z.\ acknowledges the support of NSFC (grants 12041305, 12173016) and the Program for Innovative Talents and Entrepreneurs in Jiangsu.
Y.C.\ thanks the support of NSFC grants 11773014, 11633007, and 11851305.
\end{acknowledgements}

\software{
GILDAS \citep{pety05,gildas13},
DS9 \citep{joye03,sao00},}
XSPEC \citep[vers.\ 12.9.0u,][]{arnaud96}.

\begin{appendix} 

\begin{figure}
\centering
	\includegraphics[width=0.6\textwidth]{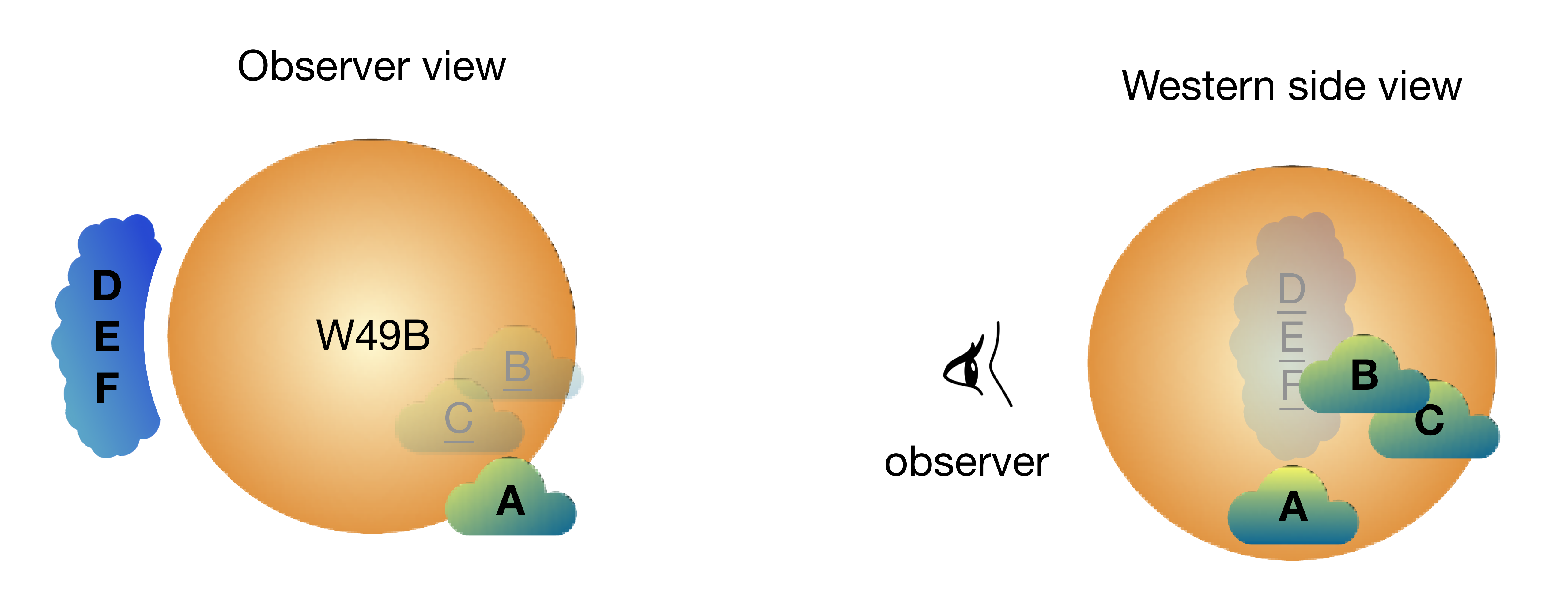}
    \caption{
    Schematic view of the interpreted positions of the six regions A--F denoted in Figure~\ref{fig:4band}. The underlined letters denote the regions behind \snr\ through the observer view or western side view.
	}
    \label{fig:cartoon}
\end{figure}

To illustrate the distributions of the six regions A--F near \snr, we provide the
schematic view in Figure~\ref{fig:cartoon}. It shows that regions B and C are located in the back side of \snr, while regions D--F are not contacted by the SNR shock. From the western side view, regions A, B, and C are in the front.

\begin{figure*}
\centering
	\includegraphics[width=0.8\textwidth]{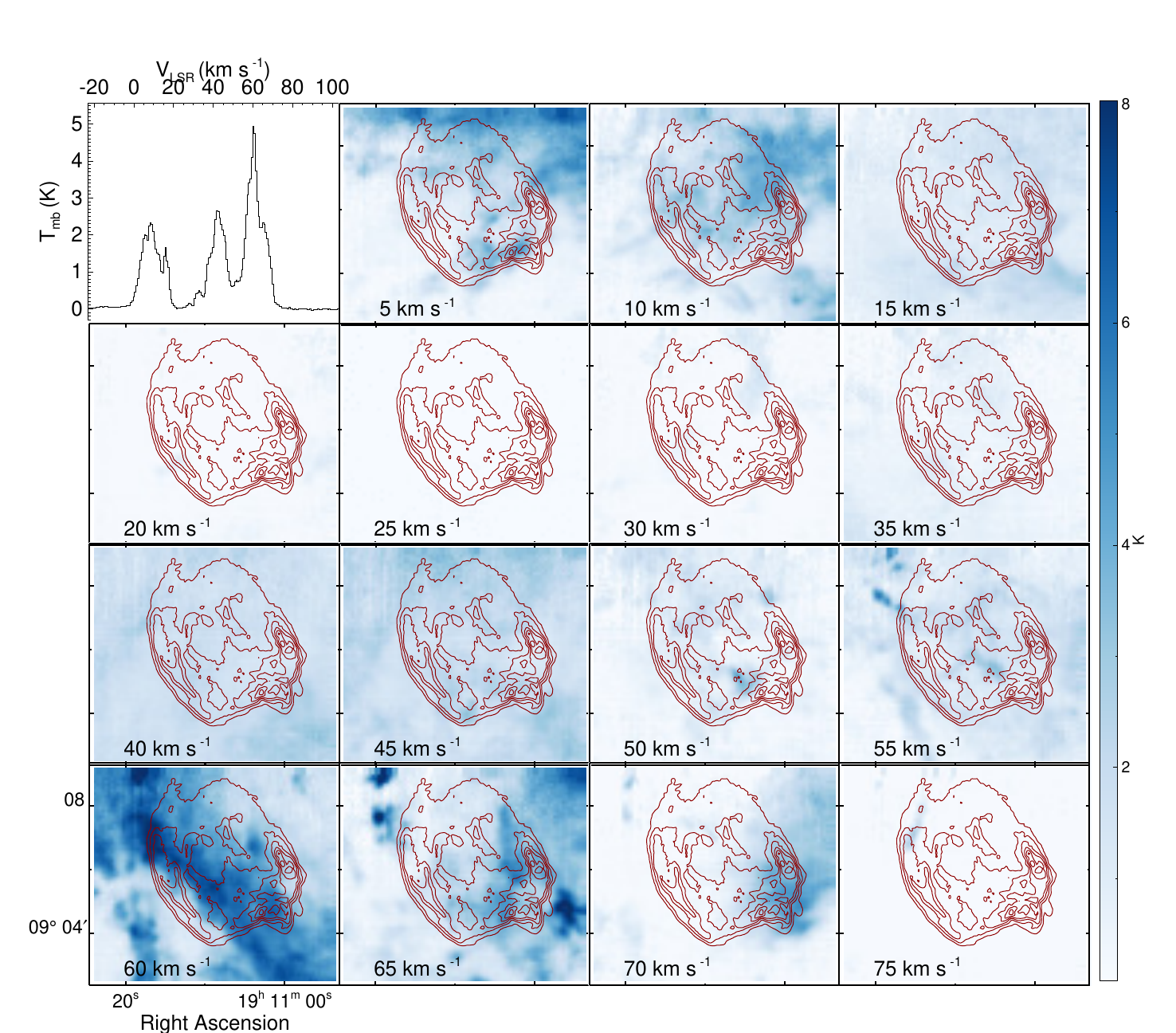}
    \caption{
    Channel map of the averaged main-beam temperature of \twCO~\Jtto\ emission with a step of 5~\kms, overlaid with contours of 327~MHz radio emission. The same contours are used for the maps in the rest of the paper. 
    The top left panel shows the spectrum averaged over the FOV.  
	}
    \label{fig:12co21_grid}
\end{figure*}

\begin{figure*}
\centering
	\includegraphics[width=0.8\textwidth]{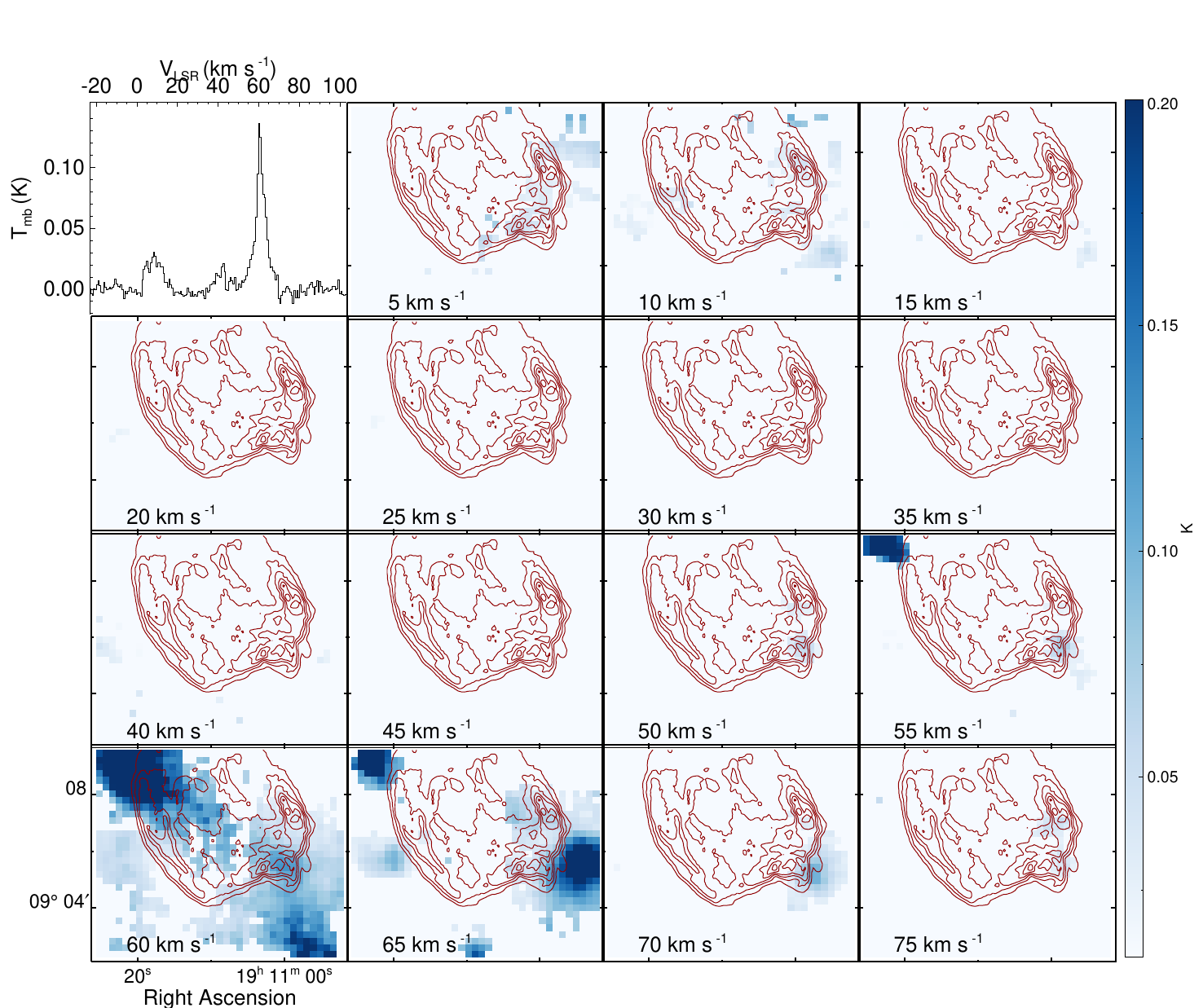}
    \caption{
      Channel map of the averaged main-beam temperature of HCO$^+$~\Jotz\ emission with a step of 5~\kms, overlaid with contours of 327~MHz radio emission. 
    The top left panel shows the spectrum averaged over the FOV. 
    Pixels with over $5\sigma$ \HCOp\ line detection are displayed.
	}
    \label{fig:hcop_grid}
\end{figure*}

Figure~\ref{fig:12co21_grid} and \ref{fig:hcop_grid} show
the distribution of \twCO~\Jtto\ and \HCOp~\Jotz\ near \snr\ in the velocity range $\VLSR=0$--80~\kms\ with a step of $5~\km\ps$.

\end{appendix}

\end{document}